\documentclass[twocolumn,english,prl,showpacs,preprintnumbers,superscriptaddress]{revtex4-1}
\usepackage[utf8]{inputenc}
\usepackage{color}
\usepackage{amsmath,amssymb}
\usepackage{amssymb}
\usepackage{graphicx}
\usepackage{textcomp}
\usepackage{dcolumn}
\usepackage{bm}
\usepackage[right]{eurosym}
\usepackage{float}
\usepackage[english]{babel}
\usepackage{blindtext}
\usepackage{babel}
\usepackage[normalem]{ulem}

\begin{document}

\title{Direct inner-shell photoionization of Xe atoms embedded in helium nanodroplets}


\author{L. Ben Ltaief} 
\affiliation{Department of Physics and Astronomy, Aarhus University, 8000 Aarhus C, Denmark}


\author{M. Shcherbinin} 
\affiliation{Department of Physics and Astronomy, Aarhus University, 8000 Aarhus C, Denmark}

\author{S. Mandal} 
\affiliation{Indian Institute of Science Education and Research, Pune 411008, India}

\author{S. Rama Krishnan} 
\affiliation{Indian Institute of Technology Madras, Chennai 600036, India}

\author{R. Richter}
\affiliation{Elettra-Sincrotrone Trieste, Basovizza, 34149 Trieste, Italy}

\author{T. Pfeifer}
\affiliation{Max-Planck-Institut f{\"u}r Kernphysik, 69117 Heidelberg, Germany}

\author{M. Mudrich} 
\affiliation{Department of Physics and Astronomy, Aarhus University, 8000 Aarhus C, Denmark}
\affiliation{Indian Institute of Technology Madras, Chennai 600036, India}
\email[E-mail me at: ]{mudrich@phys.au.dk}

\begin{abstract}
We present the first measurements of photoelectron spectra of atomic clusters embedded in superfluid helium (He) nanodroplets. Owing to the large absorption cross section of xenon (Xe) around 100 eV photon energy (4d inner-shell ionization), direct dopant photoionization exceeds charge transfer ionization via the ionized He droplets. Despite the predominant creation of Xe$^{2+}$ and Xe$^{3+}$ by subsequent Auger decay of free Xe atoms, for Xe embedded in He droplets only singly charged Xe$_k^+$, $k=1,2,3$ fragments are observed. Broad Xe$^+$ ion kinetic-energy distributions indicate Coulomb explosion of the ions due to electron transfer to the primary Auger ions from surrounding neutral atoms. The electron spectra correlated with Xe ions emitted from the He nanodroplets contain a low-energy feature and nearly unshifted Xe photolines. These results pave the way to extreme ultraviolet (XUV) and x-ray photoelectron spectroscopy of clusters and molecular complexes embedded in He nanodroplets.
\end{abstract}

\date{\today}

\maketitle

\section{Introduction}
Helium (He) nanodroplets are widely used as an ultracold matrix for spectroscopy of embedded molecules and nanostructures~\cite{Toennies:2004,Stienkemeier:2006}. The main benefits of He nanodroplets are the high resolution of absorption spectra in the infrared and visible spectral regions and the property of He droplets to efficiently form molecular aggregates that thermalize to the droplet temperature of 0.37~K. Performing spectroscopy at higher photon energies where the dopants or even the He droplets are directly ionized isn't straight forward, though; the strong interaction of photoions and electrons with the He droplet tends to massively shift and broaden the electron spectral lines and to alter the fragmentation dynamics compared to the gas phase~\cite{Mudrich:2014}. Therefore, only few photoelectron spectroscopic studies of dopants have been reported, all of which employed resonant multi-photon ionization schemes~\cite{Radcliffe:2004,Loginov:2005,Loginov:2007,Loginov:2008,Fechner:2012,Thaler:2018,Dozmorov:2018}. 

However, one-photon photoionization of doped He nanodroplets has recently turned out to be a rewarding approach for studying various types of fundamental correlated electronic decay processes such as interatomic Coulombic decay (ICD)~\cite{Cederbaum:1997,Shcherbinin:2017,Tiggesbaumker:2019,LaForge:2019,Ltaief:2019} and electron-transfer mediated decay (ETMD)~\cite{Zobeley:2001,LaForgePRL:2016,Ltaief:2020}. Although the photon energy exceeded the dopant's ionization energy $E_i$ in those studies, dopants were always ionized indirectly through the excited or ionized He. This is due to the large total absorption cross section of He nanodroplets containing thousands of He atoms ($\sim 25~$Mbarn per He atom for the dominant 1s2p$^1P$ absorption resonance at the photon energy $h\nu = 21.6~$eV) that usually largely exceeds the absorption cross section of one or a few dopant atoms or molecules. Excitations, He$^*$, and positive charges, He$^+,$ efficiently migrate through the He droplet to the dopant which is then ionized by transfer of energy or charge, respectively~\cite{Froechtenicht:1996,Wang:2008,Buchta:2013,Shcherbinin:2018,Ltaief:2019}. Large differences in the Penning ionization efficiency and the structure of the Penning electron spectra were found for dopants (alkali metals) attached to the surface of He nanodroplets compared to those immersed in the droplet interior. This was rationalized by the tendency of He$^*$ to migrate toward the droplet surface~\cite{Scheidemann:1993,Mudrich:2020}, whereas He$^+$ remains in the bulk of the droplets~\cite{Scheidemann:1993,Lehmann:1999}. Using photoelectron-photoion coincidence (PEPICO) detection, we have previously measured high Penning ionization yields for alkali metals, whereas the efficiency of Penning ionization for heavier rare gas atoms was lower than that for charge transfer ionization~\cite{Buchta:2013}. The Penning ionization electron spectra were found to feature either sharp lines reflecting the He energy levels and the dopants' $E_i$~\cite{Buchta:2013,Ltaief:2019}, a broad distribution peaked at low energies~\cite{Shcherbinin:2018}, or a combination of both~\cite{Wang:2008,Mandal:2020}.

Here, we present the first experimental study where dopant atoms attached to He nanodroplets are directly photoionized and electron and ion spectra are recorded. This is achieved using Xe as a dopant and setting $h\nu\sim 100$~eV where Xe features a pronounced maximum of the 4d-shell ionization cross section, whereas the absorption cross section of He is down by a factor $\sim 1/20$ compared to the value near $E_i^\mathrm{He}$. Photoions from free atoms in the gas phase are mostly produced in doubly and triply charged states as a result of normal or cascaded Auger decay, respectively. In contrast, from doped He droplets mostly singly charged Xe$^+$ as well as small Xe$_n^+$ clusters are emitted. This points at efficient partial neutralization of highly charged cations in He nanodroplets by electron transfer to the dopant photoion from neutral dopant atoms surrounding it. Electron spectra exhibit sharp unshifted photolines from the embedded Xe clusters as well as a pronounced low-energy distribution indicative for electron-He scattering and electron localization.

\section{Experimental}
The experiments are performed using a He nanodroplet apparatus combined with a velocity-map imaging photoelectron-photoion coincidence (VMI-PEPICO) spectrometer installed at the GasPhase beamline of Elettra-Sincrotrone Trieste, Italy. The apparatus has been described in detail elsewhere~\cite{Buchta:2013,BuchtaJCP:2013}. Briefly, a beam of He nanodroplets is produced by continuously expanding pressurized He (50~bar) of high purity out of a cold nozzle (14~K) with a diameter of 5~$\mu$m into vacuum, resulting in a mean droplet size of $\bar{N}_\mathrm{He} = 2.3\times 10^4$ He atoms per droplet. The He droplets were doped with Xe atoms by leaking Xe gas into a doping gas cell of length 30~mm. The measurements presented in this paper were done at a Xe pressure in the doping cell of $4.3\times 10^{-4}~$mbar. This corresponds to a mean number of 24 Xe dopants per He droplet. A mechanical beam chopper at the entrance of the doping chamber is used for discriminating droplet-beam correlated signals from the background. 
In the detector chamber, the He droplet beam crosses the synchrotron beam at the center of the VMI-PEPICO detector at right angles. By detecting either electrons or ions with the VMI detector in coincidence with the corresponding particles of opposite charge on the TOF detector, we obtain either ion mass-correlated electron images or mass-selected ion images. Kinetic-energy distributions of electrons and ions are obtained by Abel inversion of the images~\cite{Dick:2014}. The energy resolution of the electron spectra obtained in this way is $\Delta E/E=6$\%.

\section{Results and discussion}
In our previous PEPICO study of Ar-doped He nanodroplets it appeared that heavier rare gas atoms solvated in the droplet interior are inefficiently Penning ionized through excited He~\cite{Buchta:2013}. In contrast, Wang~\textit{et al.} had previously reported well-resolved Penning electron spectra of Kr and Xe indicating that Penning ionization of Kr and Xe embedded in He nanodroplets is quite efficient~\cite{Wang:2008}. 

\begin{figure}
	\center
	\includegraphics[width=1.0\columnwidth]{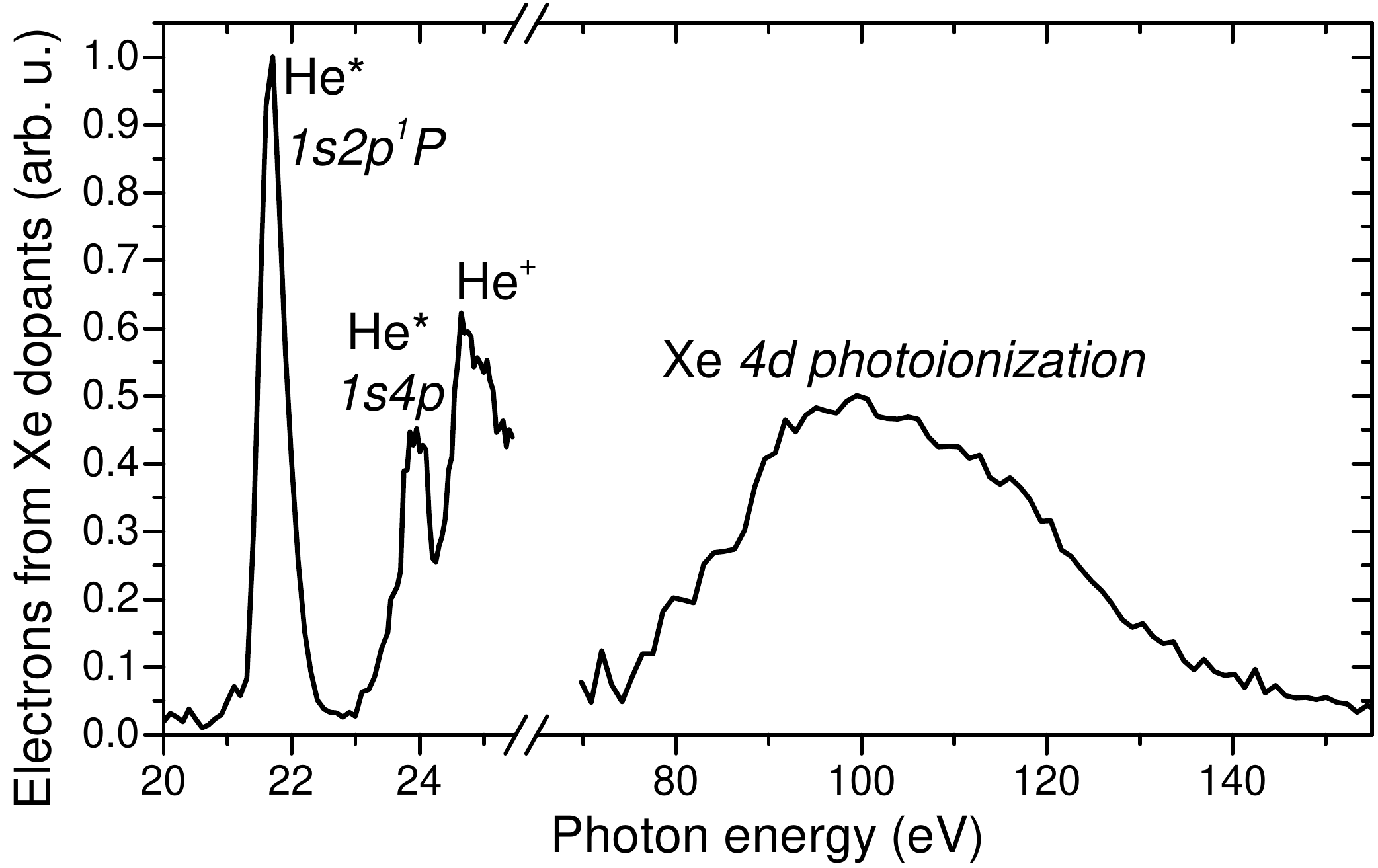}\caption{\label{fig:Xe_espec} Electron yield spectrum due to Xe atoms embedded in He nanodroplets as a function of the photon energy. The signal below the ionization energy of He (to the left of the break in the $h\nu$-axis) is due to indirect ionization of Xe through excited and autoionized He, the part at $h\nu > 70~$eV is mostly due to direct photoionization of the Xe dopants.}
\end{figure}

To resolve these conflicting findings, we record the total electron yield of Xe-doped He nanodroplets in the wide photon energy range $h\nu =$20-160~eV. Those electrons emitted from Xe dopants embedded in He droplets are extracted from the data by first subtracting from the total electron signal (chopper open) those electrons emitted by ionization of the background gas (chopper closed). Then, we subtract from the measurement done with Xe doping on a reference measurement where the Xe doping was turned off. The resulting electron yield spectrum is shown in Fig.~\ref{fig:Xe_espec}.

In the range $h\nu < E_i^\mathrm{He}$, the electron yield closely follows the absorption spectrum of pure He nanodroplets, which is dominated by the 1s2p$^1P$ absorption resonance of He nanodroplets peaked at $h\nu = 21.6~$eV~\cite{Joppien:1993}. 
We conclude that Xe is indeed efficiently Penning ionized by excited He nanodroplets, similar to our previous finding for alkali and alkaline-earth metals~\cite{Buchta:2013,Mudrich:2014,LaForgePRL:2016}. The reason why we previously measured much lower Penning ionization signals from rare-gas dopants than from the metals was that rare-gas Penning ions formed inside the He droplets tend to remain bound to the droplets even if the Penning electrons are emitted. As we detected electrons and ions in coincidence, both escaped detection. Nevertheless, our conclusion that the surface location of alkali atoms facilitates He droplet Penning ionization remains true. The proportion of Penning ionization signals measured at $h\nu = 21.6~$eV versus dopant ionizations by charge transfer at $h\nu > 24.6~$eV is about 5 times higher for alkali metals than for Xe~\cite{Buchta:2013}. Besides, it has been shown by electron impact ionization that small alkali clusters residing at the droplet surface are more efficiently Penning ionized than large alkali clusters which sink into the droplet interior~\cite{Lan:2011,Lan:2012}.

At higher XUV photon energies $h\nu>70~$eV, the yield of electrons is lower than that measured at the He 1s2p$^1P$ resonance but clearly shows a broad maximum centered around $h\nu\sim 100~$eV. In this range of $h\nu$ the electron yield closely follows the absorption cross section of Xe atoms which is dominated by a maximum of the 4d-subshell photoionization cross section, also called `giant resonance'~\cite{Ederer:1964,Becker:1989}. This is a clear indication that now the detected electrons are mostly emitted by the Xe dopants. Photoionization of the He droplets followed by charge transfer ionization of the Xe dopants, which is the dominant dopant-ionization mechanism near $E_i^\mathrm{He}$, contributes to a lesser extent. This is due to the large difference in absorption cross sections of Xe (23.6~Mbarn) and He (0.52~Mbarn) at $h\nu = 90~$eV~\cite{Samson:2002}. Given the droplet size of about $2.3\times 10^4$ He atoms and the estimated Xe dopant clusters size of 24 Xe atoms, we obtain a ratio of the efficiencies of direct photoionization of embedded Xe vs. indirect charge transfer ionization of about 7 assuming a charge transfer ionization probability of the Xe cluster of 1\,\%~\cite{Callicoatt:1996,Ruchti:2000}. This value is in good agreement with the signal contrast from on-resonant ($h\nu\sim 100~$eV) Xe photoionization with respect to the off-resonant ($h\nu\sim 150~$eV) background measured here (Fig.~1). Thus, we have demonstrated for the first time that direct one-photon ionization of dopants embedded in He nanodroplets is possible, at least at high XUV photon energies $h\nu\gg E_i^\mathrm{He}$ where He nanodroplets are nearly transparent.  

\begin{figure}
	\center
	\includegraphics[width=1.0\columnwidth]{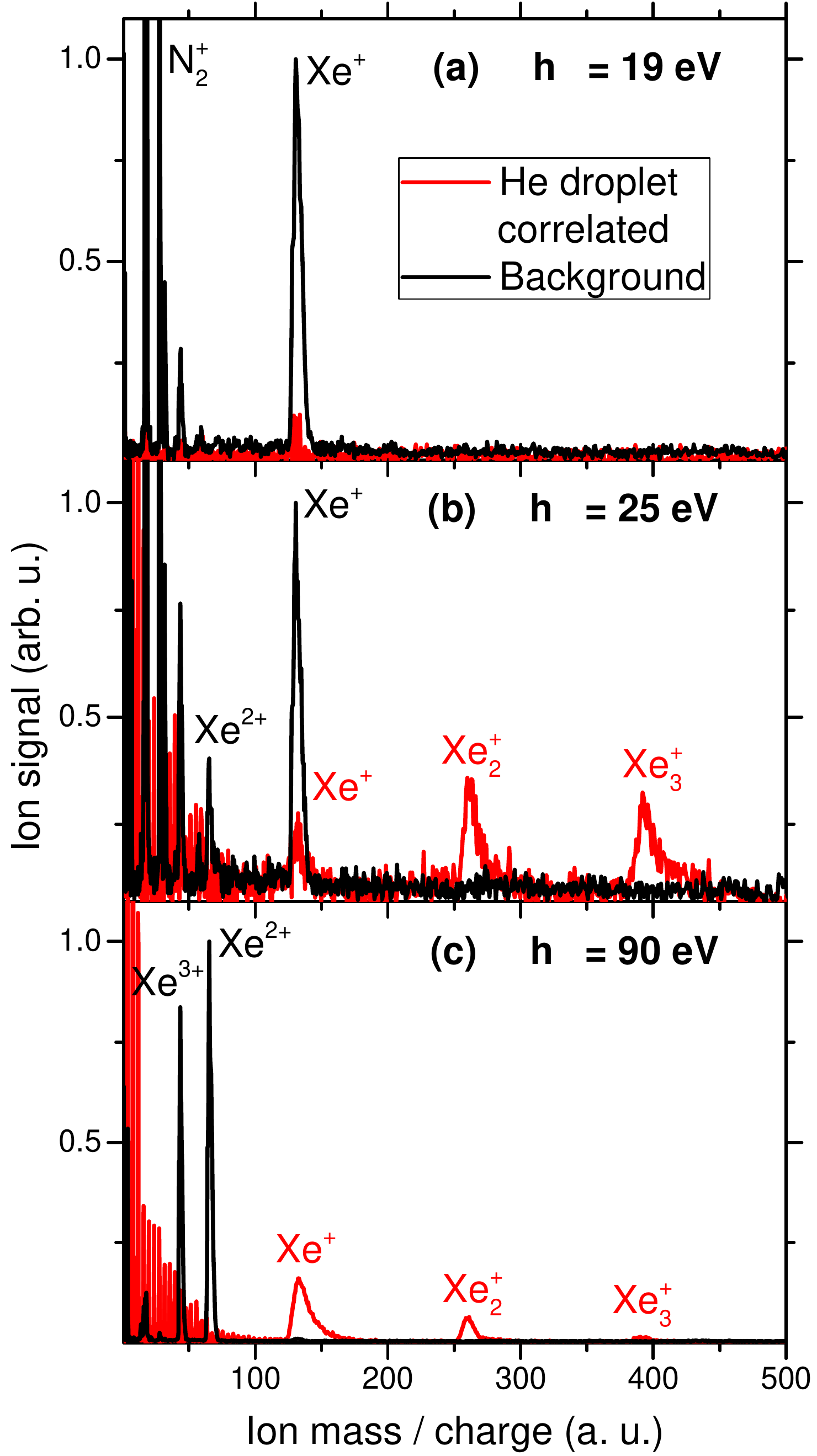}\caption{\label{fig:Xe_massspec} Mass spectra of Xe-doped He nanodroplets recorded at different photon energies. The ion signal correlated to the doped He nanodroplets (red line) is discriminated from the background (black line) through a mechanical chopper that interrupts the He nanodroplet beam.}
\end{figure}
When Xe is 4d-subshell ionized, a cascaded Auger decay takes place resulting in multiply charged Xe ions~\cite{Becker:1989}. Fig~\ref{fig:Xe_massspec} (c) shows typical mass spectra recorded for Xe atoms in the background gas (black line) and for Xe embedded in He nanodroplets (red line) at $h\nu =90~$eV. Clearly, the dominant charge states from Auger decay are Xe$^{2+}$ and Xe$^{3+}$, whereas Xe$^{+}$ is hardly visible when normalizing the ion signal scale to the Xe$^{2+}$ peak. Note that the abundances of the highly charged Xe$^{2+}$ and Xe$^{3+}$ ions are likely enhanced compared to Xe$^{+}$ and in particular to Xe$_{2}^+$ and Xe$_{3}^+$ due to a higher detection sensitivity. In contrast, just below and above $E_i^\mathrm{He}$ (Fig~\ref{fig:Xe_massspec} (a) and (b)), Xe$^{+}$ is by far the most abundant product. The small contribution of Xe$^{2+}$ in the mass spectrum at $h\nu = 25~$eV is likely due to one-photon double ionization of Xe by second-order synchrotron radiation which is quite abundant at that photon energy.

The He droplet-correlated Xe$^{+}$ signal at $h\nu =19~\mathrm{eV} < E_i^\mathrm{He}$ (red line in Fig~\ref{fig:Xe_massspec} (a)) is nearly absent. The small Xe$^{+}$ peak likely stems from imperfect discrimination of the He droplet-correlated signal from the background. As the He droplets are neither excited nor ionized at that photon energy, indirect ionization of Xe through the He is suppressed and direct photoionization of Xe dopants does not significantly contribute to the ion signal. At $h\nu =25~\mathrm{eV} > E_i^\mathrm{He}$ (red line in Fig~\ref{fig:Xe_massspec} (b)), however, He droplet-correlated Xe ions are present mostly as small Xe$_k^+$ clusters, $k=1,~2,~3$. The fragmentation of these small Xe dopant clusters is due to charge transfer ionization through the ionized He nanodroplets, as it has been discussed in Ref.~\cite{Ruchti:2000}. 

One main result of this work is that at $h\nu =90~\mathrm{eV}$, the He droplet-correlated Xe$^{+}$ mass spectrum (red line in Fig~\ref{fig:Xe_massspec} (c)) again consists of small singly charged Xe$_k^+$ clusters, although Xe ionization is mainly due to direct Xe 4d-ionization which predominantly creates Xe$^{2+}$ and Xe$^{3+}$. Thus, multiply charged ions in He nanodroplets are very efficiently partly neutralized. Electron transfer from neighboring neutral atoms to the highly charged ions can occur on a subfemtosecond time scale, \textit{i.~e.} faster than the Auger process~\cite{Gnodtke:2009}. Partial neutralization of doubly charged metals in He nanodroplets created by electron-transfer mediated decay was observed previously~\cite{LaForgePRL:2016}. Even pure Xe clusters irradiated by soft and hard x-rays have recently been found to efficiently quench high charge states created by Auger ionization~\cite{oelze:2017,Kumagai:2018}. 
This was interpreted by electron-ion recombination which is a common process in expanding nanoplasmas~\cite{oelze:2017}. Here, we show that electron transfer to multiply charged ions is highly efficient even in the absence of a nanoplasma. Likely, in~\cite{oelze:2017} also electron transfer from neutral Xe contributed to the measured highly abundant Xe$^{+}$ signals~\cite{oelze:2017}. In our experiment (Fig~\ref{fig:Xe_massspec} (c)), where the He nanodroplets contain small Xe clusters, likely electron transfer between Xe atoms is the main mechanism. We note that even when we reduced the number of doped Xe atoms down to the detection limit of Xe ions from He droplets in an attempt to dope the He droplets by single Xe atoms, no Xe$^{2+}$ were present although the latter cannot be neutralized by He as $E_i^\mathrm{Xe^+}<E_i^\mathrm{He}$. The likely reason is that a multiply charged cation is strongly bound to a He droplet by forming a so-called snowball complex~\cite{Atkins:1959} and thus evades its detection. Likewise, no experimental evidence for charge transfer to the highly charged Xe$^{3+}$ from surrounding neutral He atoms was found in this work, although it would be energetically allowed. The possible charge transfer processes that can occur between Xe atoms embedded inside He nanodroplets are as follows:\\

\noindent
$\mathrm{Xe}_{n}\mathrm{He}_{N}+h\nu\rightarrow\mathrm{Xe}_{n}^{2+}\mathrm{He}_{N}+e_{Aug}+e_{ph}$
\begin{itemize}
    \item[$\rightarrow$]$\mathrm{Xe}_{n-k-1}\mathrm{He}_{N}+\mathrm{Xe}^{+}+\mathrm{Xe}_k^{+}+e_{Aug}+e_{ph}$, where $k=1,2,3$.
\end{itemize}
Or, $\mathrm{Xe}_{n}\mathrm{He}_{N}+h\nu\rightarrow\mathrm{Xe}_{n}^{3+}$He$_{N}+2e_{Aug}+e_{ph}$ 
\begin{itemize}
    \item[$\rightarrow$]$\mathrm{Xe}_{n-k-1}$He$_{N}+\mathrm{Xe}^{2+}+\mathrm{Xe}_k^{+}+2e_{Aug}+e_{ph}$
    \item[$\rightarrow$]$\mathrm{Xe}_{n-3}\mathrm{He}_{N}+\mathrm{Xe}^{+}+\mathrm{Xe}^{+}+\mathrm{Xe}^{+}+2e_{Aug}+e_{ph}$.
\end{itemize}

When we compare the Xe$^{+}_k$ distribution measured by charge transfer ionization (Fig~\ref{fig:Xe_massspec} (b)) with the one by Auger ionization in conjunction with electron transfer (Fig~\ref{fig:Xe_massspec} (c)), we note that the Xe$^+$ peak is higher and broader in the latter case. This is likely due to Coulomb explosion of the two or three Xe$^+$ ions formed from Xe$^{2+}$ or Xe$^{3+}$ by electron transfer, respectively. When two ions are formed with substantial kinetic energy, they are less prone to being trapped by the He droplet and are therefore detected with higher probability~\cite{LaForgePRL:2016,LaForge:2019}. 

\begin{figure}
	\center
	\includegraphics[width=1.0\columnwidth]{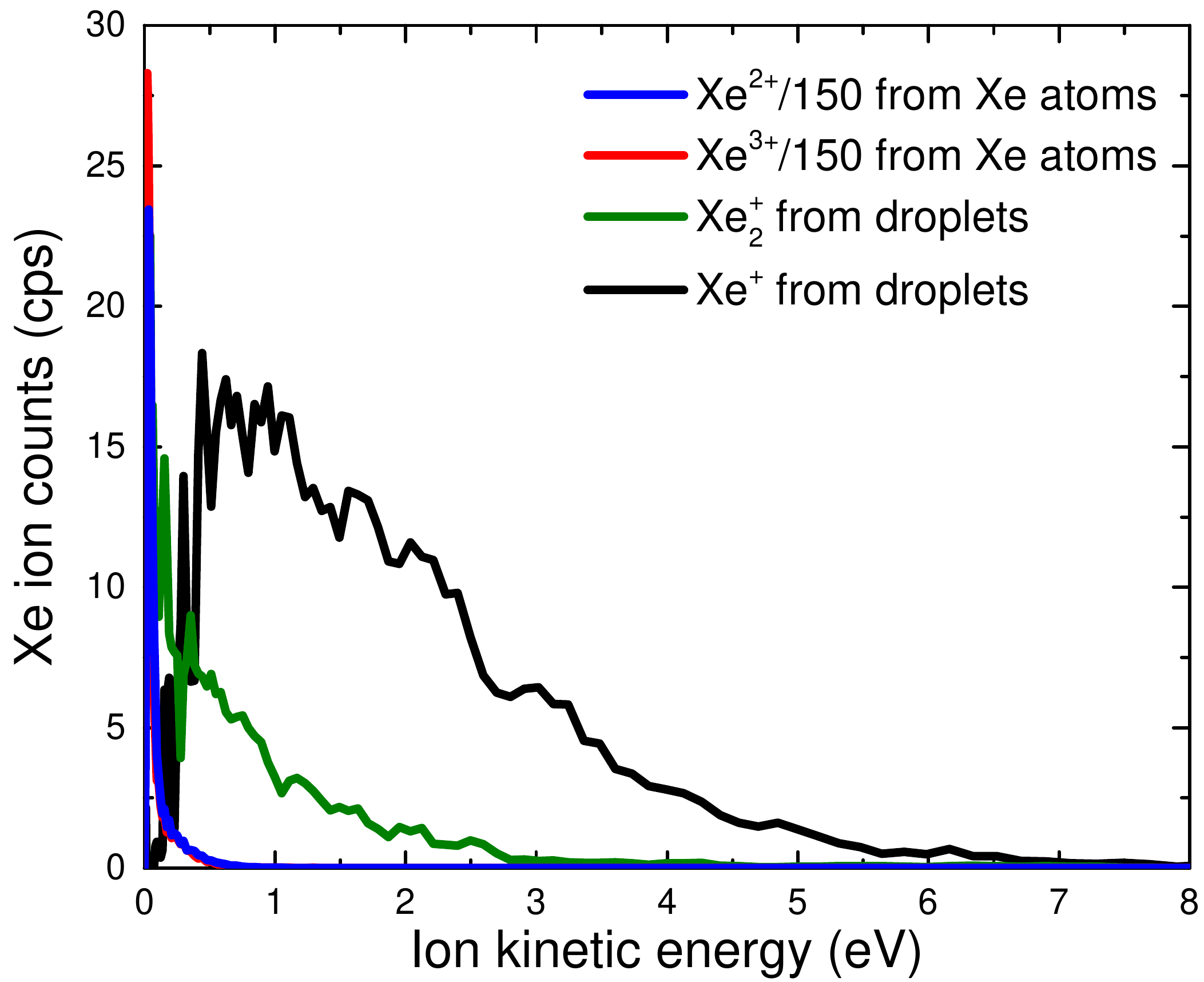}\caption{\label{fig:Ionspec} Kinetic energy distributions of Xe$^{+}$ and Xe$_{2}^+$ ions created from Xe doped He droplets in comparison with Xe$^{2+}$ and Xe$^{3+}$ Auger ions from free Xe atoms. All ion spectra are measured at $h\nu =90~\mathrm{eV}$.}
\end{figure}

Further evidence for the formation of Xe$^+$ by Coulomb explosion is obtained from directly measuring the ion kinetic energy by velocity-map imaging the Xe$^+$ on a position sensitive detector in the ion-imaging mode. The Xe$^+$ and Xe$_{2}^+$ ion kinetic energy distributions inferred from ion images are displayed in Fig. 3 in comparison with the ion kinetic energy distributions for Xe$^{2+}$ and Xe$^{3+}$ from free Xe atoms. The Xe$^{+}$ ion spectrum consists of a broad feature that peaks around 1.5~eV and extends up to 6.5~eV, whereas the Xe$^{2+}$ and Xe$^{3+}$ spectra both exhibit only a very narrow peak at 0~eV. The width of these peaks reflects the experimental resolution. 

For Xe dimers, the bond length is $R=4.36$~Å~\cite{tang:2003}.
Assuming instantaneous formation of an ion pair Xe$^+$ + Xe$^+$ by Auger decay and electron transfer, the kinetic energy release (KER) due to Coulomb explosion according to the repulsive Coulomb potential $e^{2}/(4\pi\varepsilon_{0} R)$ is estimated to 3.3~eV. Thus, each Xe$^{+}$ ion acquires a kinetic energy of 1.7~eV, which is in good agreement with the maximum of the measured kinetic energy distribution. For larger Xe clusters, the KER is expected to be higher since the interatomic distance between two nearest-neighbor atoms is slightly shorter as it approaches the bond length in bulk Xe, $4.26$~Å~\cite{hermann:2009}. 

The tail in the kinetic energy distribution extending up to 6.5~eV in the Xe$^+$ is likely due to Coulomb explosion of three Xe$^{+}$ ions after creation of one Xe$^{3+}$ by Auger decay followed by electron charge transfer from two neighboring Xe atoms. The kinetic energy of the Xe$^{+}$ for the Coulomb explosion of a Xe trimer system are expected to range between 3.3~eV and 4.5~eV depending on the initial configuration. Furthermore, when Coulomb explosion occurs in a larger Xe$_{k}$ cluster where one charge is localized on one Xe atom and the other is localized on the remaining cluster Xe$_{k-1}$, the Xe$^+$ acquires a kinetic energy up to the full KER in the limit of a very large Xe$_{k-1}^+$. This kinematic effect adds to the asymmetric broadening of the Xe$^+$ kinetic energy distribution towards higher energies.

The Xe$_{2}^+$ ion spectrum shows a bimodal distribution with a trailing edge (0.5-4~eV) that resembles the one of the Xe$^+$ ion spectrum (1-6~eV) but scaled down to lower energy. Again, this may be due to the kinematic effect, from which we expect a factor of 2 lower energy of Xe$_{2}^+$ than Xe$^+$ for the case that Coulomb explosion occurs from the Xe$_{3}^+$ system. The peak at $< 0.1$~eV seen in the Xe$_{2}^+$ spectrum might be related to a non-thermal ejection process that occurs for vibrationally excited molecular ions~\cite{smolarek:2010}, assuming that part of the Coulomb exploding Xe$_{2}^+$ are fully decelerated by collisions with surrounding Xe and He atoms in the droplets prior to ejection.

\begin{figure}
	\center
	\includegraphics[width=1.0\columnwidth]{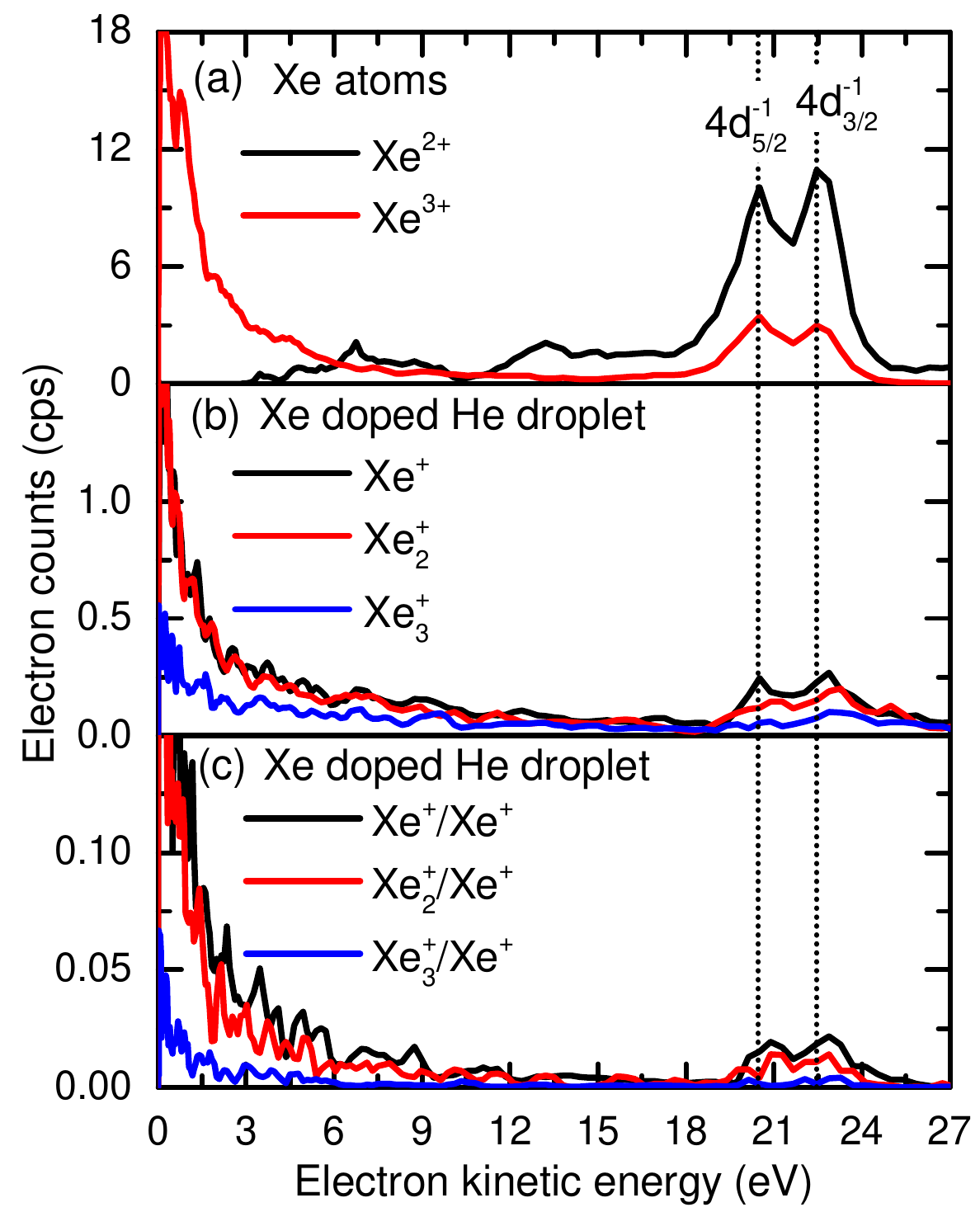}\caption{\label{fig:Xe_PES} Electron spectra recorded for Xe atoms (a) and for He droplets doped with Xe atoms ((b) and (c)) at photon energy $h\nu =90~\mathrm{eV}$. Black and red spectrum in (a) are recorded in coincidence with Xe$^{2+}$ and Xe$^{3+}$, respectively. The spectra denoted in black, red and blue in (b) recorded in coincidence with Xe$^{+}$, Xe$_{2}^{+}$ and Xe$_{2}^{+}$, respectively. The black, red and blue curves in (c) show electron spectra measured in double coincidence with Xe$^{+}$/Xe$^{+}$, Xe$^{+}$/Xe$_{2}^{+}$ and Xe$^{+}$/Xe$_{3}^{+}$, respectively. The vertical dashed lines show the energy positions of the atomic 4d$_{3/2}^{-1}$ and 4d$_{5/2}^{-1}$ lines according to ~\cite{lablanquie:2002}.}
\end{figure}

Fig. 4 shows electron spectra for Xe atoms measured in coincidence with atomic Xe$^{2+}$ and Xe$^{3+}$ ions (panel (a)) as well as those measured in coincidence with Xe$^{+}$, Xe$_{2}^{+}$ and Xe$_{3}^{+}$ emitted from He droplets (panels (b) and (c)).
The two energy-resolved 4d$_{5/2}^{-1}$ and 4d$_{5/2}^{-1}$ lines seen in the atomic PES spectra for Xe$^{2+}$ and Xe$^{3+}$ are also present in the He droplet-correlated electron spectra (Panel (b)). This is another important result of this work. The low-energy part of the He droplet-correlated electron spectra contains a pronounced feature that resembles the low-energy Auger electron spectrum of the atomic Xe PES measured in coincidence with Xe$^{3+}$. 
However, the fraction of electrons of low-kinetic energy vs. the photolines is larger for the He droplet-correlated PES than for the charge-state averaged atomic PES by a factor of 2.5. This implies that in He droplets, part of the photoelectrons are slowed to low kinetic energies by electron-He scattering.

Following inner shell ionization of Xe clusters inside He droplets, ICD or ETMD driven by Auger decay could be competing processes to the local atomic Auger decay~\cite{morishita:2006, ueda:2007, ueda:2008, sakai:2011}. These processes typically generate slow electrons which may add to the low-energy part of the He droplet-correlated electron spectra observed here. Direct experimental evidence would be the detection of Xe$^{2+}$ in coincidence with Xe$^{+}$. However, the Xe$^{2+}$ signal is quenched by the rapid charge transfer occurring in clusters quenches. Direct inner-shell ICD could also take place, but its observation is hampered by the low branching ratio with respect to local Auger decay. Core-level ICD in rare gas dimers and clusters has recently been found to contribute by only 0.26-0.8~$\%$ of the local Auger decay~\cite{Liu:2020, hans:2020}. 

Overall the resulting electron spectra, which are characterized by sharp atomic peaks at high energy and a broad tail that rises towards zero-kinetic energy, have a similar structure as previously measured Penning ionization electron spectra of dopants in He nanodroplets~\cite{Wang:2008,Mandal:2020}. The presence of these two features in the spectra may be related to the dopants occupying different states close to the droplet surface or deep inside the droplets.

Further evidence for the presence of sharp atomic peaks in the He droplet-correlated Xe electron spectra is obtained from the electron spectra for electron-ion-ion triple coincidence events, shown in panel (c). Interestingly, the 4d$_{3/2,\,5/2}^{-1}$ photolines in the electron-Xe$^+$-Xe$^+$ triple-coincidence spectra as well as in the electron-Xe$_{2,\,3}^+$ double-coincidence spectra are slightly shifted towards higher kinetic energies as compared to those of Xe atoms. Fig.~5 clearly shows this energy shift on an enlarged scale, which amounts to about 0.2~eV (4d$_{5/2}^{-1}$) and 0.4~eV (4d$_{3/2}^{-1}$) for electrons measured in coincidence with Xe$^{+}$ and to about 0.4~eV (4d$_{5/2}^{-1}$) and 0.7~eV (4d$_{3/2}^{-1}$) for electrons measured in coincidence with Xe$_{2}^{+}$. The red lines depict Lorentzian fit functions whose widths were set to the resolution of our spectrometer. These energy shifts are consistent with the shifts of inner-shell levels measured in highly resolved electron spectra of free Xe clusters~\cite{bjorneholm:2009, tchaplyguine:2004}. The latter exhibited well resolved peaks assigned to surface and bulk atoms that were shifted by 0.8 and 1.2~eV, respectively, with respect to the free atomic 4d$_{3/2}^{-1}$ and 4d$_{5/2}^{-1}$ photolines. Thus, within the limited energy resolution of the current experiment, the He droplet environment does not seem to induce additional shifts and broadening of the photolines of the embedded Xe clusters. The fact that the photolines measured in coincidence with Xe$^+$ are nearly unshifted (Fig.~4 b)) may indicate that Xe$^+$ atomic fragments tend to be emitted from 4d-ionized Xe$_2$ or small Xe clusters whose inner-shell electron spectra are only weakly perturbed, whereas larger Xe clusters, which feature more strongly perturbed electron spectra, fragment more likely into Xe$_2^+$ and Xe$_3^+$.
\begin{figure}
	\center
	\includegraphics[width=0.8\columnwidth]{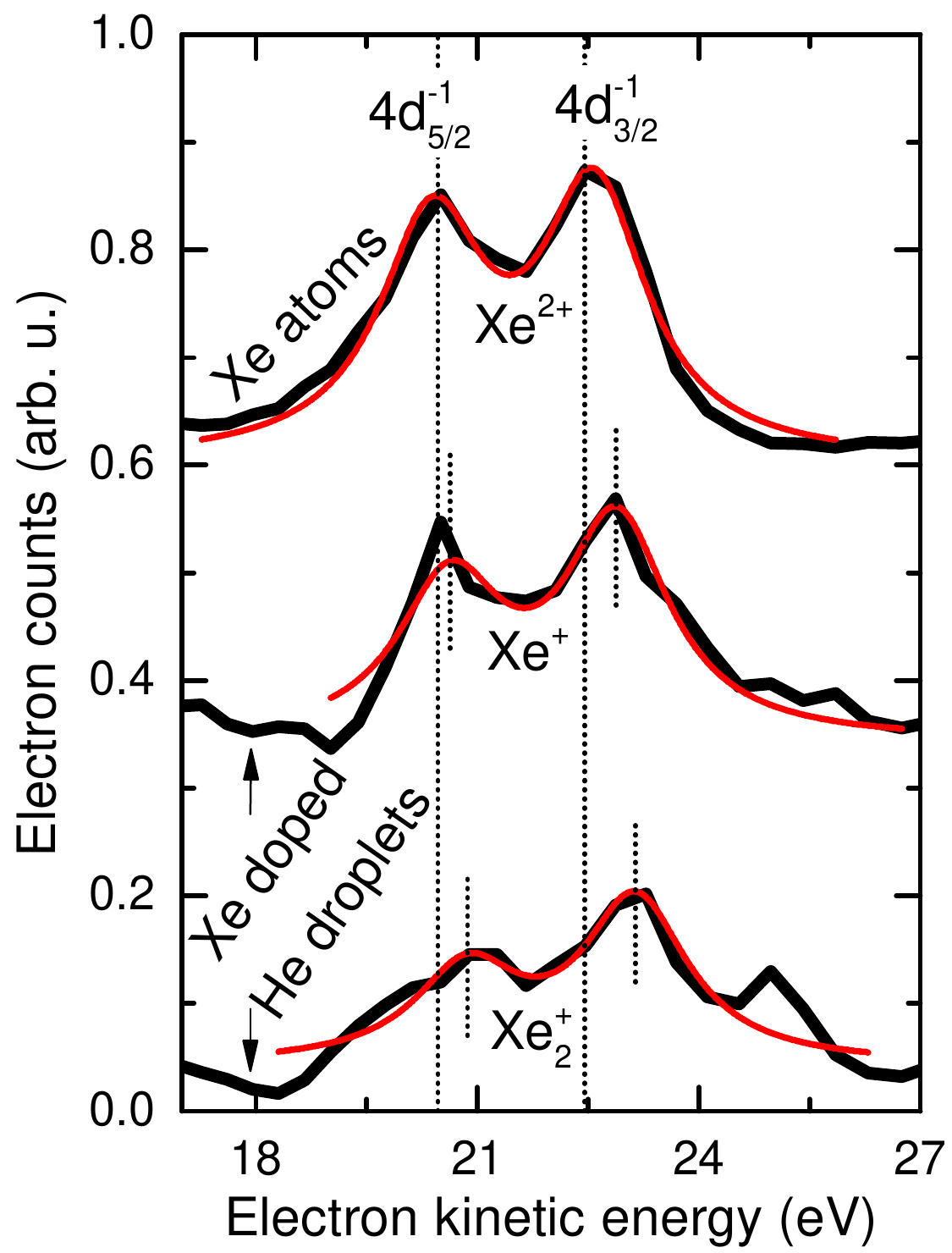}\caption{\label{fig:Xe_PES_fit} Close-ups of the 4d$_{5/2}^{-1}$ and 4d$_{5/2}^{-1}$ photolines in case of free Xe atoms and Xe clusters formed inside He nanodroplets and irradiated with an XUV photons of energy $h\nu =90~\mathrm{eV}$. The solid curves depicted in red are Lorentzian fit functions to determine the energy positions of the 4d$_{5/2}^{-1}$ and 4d$_{5/2}^{-1}$ photolines in each electron spectrum.}
\end{figure}
\section{Conclusion}
We have reported the first experimental evidence of direct one-photon ionization of dopants embedded in He nanodroplets. We exploit the large absorption cross section for 4d inner-shell ionization of Xe at a photon energy around $h\nu = 100~$eV where He has a low absorption cross section. For Xe clusters formed inside He nanodroplets, multiply charged Xe atoms created by Auger decay are efficiently partially neutralized into singly charged Xe$^+_k$, $k=1,2,3$, clusters by electron transfer from surrounding neutral Xe and He atoms. Subsequent Coulomb explosion generates Xe$^+$ ions with up to 6.5~eV of kinetic energy. The electron spectra of droplet-bound Xe clusters feature both an enhanced low-energy component indicative for electron-He scattering, and nearly unshifted 4d photolines.

These results demonstrate that photoelectron spectroscopy of clusters embedded in He nanodroplets in the XUV range is possible. Very likely, the same holds for x-rays, provided a sufficiently sensitive detection scheme is used that copes with the low target density. This paves the way to x-ray photoelectron spectroscopy (XPS) of unconventional atomic clusters and molecular complexes, which can form in He nanodroplets owing to their unique quantum fluid properties~\cite{Higgins:1996,NautaScience:1999,NautaScience:2000,Przystawik:2008}. Furthermore, the selective multiple ionization of dopants in He nanodroplets is an efficient mechanism for igniting a nanoplasma~\cite{Schomas:2020}. Probing the dynamics of nanoplasmas by ultrashort XUV and x-ray pulses has attracted considerable attention in the free-electron laser science community~\cite{Wabnitz:2002,Hoener:2008,Gorkhover:2016,Kumagai:2018}.

\section*{Acknowledgement}
M.M. and L.B.L. acknowledge financial support by Deutsche Forschungsgemeinschaft (DFG, German Research Foundation, projects MU 2347/10-1 and BE 6788/1-1) and by the Carlsberg Foundation. SRK thanks DST and MHRD, Govt of India, through the IMPRINT programmes, and the Max Planck Society. M.M. and S.R.K. gratefully acknowledge funding from the SPARC Programme, MHRD, India. The research leading to this result has been supported by the project CALIPSOplus under grant agreement 730872 from the EU Framework Programme for Research and Innovation HORIZON 2020.


\begin{thebibliography}{60}%
\makeatletter
\providecommand \@ifxundefined [1]{%
 \@ifx{#1\undefined}
}%
\providecommand \@ifnum [1]{%
 \ifnum #1\expandafter \@firstoftwo
 \else \expandafter \@secondoftwo
 \fi
}%
\providecommand \@ifx [1]{%
 \ifx #1\expandafter \@firstoftwo
 \else \expandafter \@secondoftwo
 \fi
}%
\providecommand \natexlab [1]{#1}%
\providecommand \enquote  [1]{``#1''}%
\providecommand \bibnamefont  [1]{#1}%
\providecommand \bibfnamefont [1]{#1}%
\providecommand \citenamefont [1]{#1}%
\providecommand \href@noop [0]{\@secondoftwo}%
\providecommand \href [0]{\begingroup \@sanitize@url \@href}%
\providecommand \@href[1]{\@@startlink{#1}\@@href}%
\providecommand \@@href[1]{\endgroup#1\@@endlink}%
\providecommand \@sanitize@url [0]{\catcode `\\12\catcode `\$12\catcode
  `\&12\catcode `\#12\catcode `\^12\catcode `\_12\catcode `\%12\relax}%
\providecommand \@@startlink[1]{}%
\providecommand \@@endlink[0]{}%
\providecommand \url  [0]{\begingroup\@sanitize@url \@url }%
\providecommand \@url [1]{\endgroup\@href {#1}{\urlprefix }}%
\providecommand \urlprefix  [0]{URL }%
\providecommand \Eprint [0]{\href }%
\providecommand \doibase [0]{https://doi.org/}%
\providecommand \selectlanguage [0]{\@gobble}%
\providecommand \bibinfo  [0]{\@secondoftwo}%
\providecommand \bibfield  [0]{\@secondoftwo}%
\providecommand \translation [1]{[#1]}%
\providecommand \BibitemOpen [0]{}%
\providecommand \bibitemStop [0]{}%
\providecommand \bibitemNoStop [0]{.\EOS\space}%
\providecommand \EOS [0]{\spacefactor3000\relax}%
\providecommand \BibitemShut  [1]{\csname bibitem#1\endcsname}%
\let\auto@bib@innerbib\@empty
\bibitem [{\citenamefont {Toennies}\ and\ \citenamefont
  {Vilesov}(2004)}]{Toennies:2004}%
  \BibitemOpen
  \bibfield  {author} {\bibinfo {author} {\bibfnamefont {J.~P.}\ \bibnamefont
  {Toennies}}\ and\ \bibinfo {author} {\bibfnamefont {A.~F.}\ \bibnamefont
  {Vilesov}},\ }\href@noop {} {\bibfield  {journal} {\bibinfo  {journal}
  {Angew. Chem. Int. Ed.}\ }\textbf {\bibinfo {volume} {43}},\ \bibinfo {pages}
  {2622} (\bibinfo {year} {2004})}\BibitemShut {NoStop}%
\bibitem [{\citenamefont {Stienkemeier}\ and\ \citenamefont
  {Lehmann}(2006)}]{Stienkemeier:2006}%
  \BibitemOpen
  \bibfield  {author} {\bibinfo {author} {\bibfnamefont {F.}~\bibnamefont
  {Stienkemeier}}\ and\ \bibinfo {author} {\bibfnamefont {K.}~\bibnamefont
  {Lehmann}},\ }\bibfield  {title} {\bibinfo {title} {Spectroscopy and dynamics
  in helium nanodroplets},\ }\href@noop {} {\bibfield  {journal} {\bibinfo
  {journal} {J.~Phys.~B}\ }\textbf {\bibinfo {volume} {39}},\ \bibinfo {pages}
  {R127} (\bibinfo {year} {2006})}\BibitemShut {NoStop}%
\bibitem [{\citenamefont {Mudrich}\ and\ \citenamefont
  {Stienkemeier}(2014)}]{Mudrich:2014}%
  \BibitemOpen
  \bibfield  {author} {\bibinfo {author} {\bibfnamefont {M.}~\bibnamefont
  {Mudrich}}\ and\ \bibinfo {author} {\bibfnamefont {F.}~\bibnamefont
  {Stienkemeier}},\ }\bibfield  {title} {\bibinfo {title} {Photoionisaton of
  pure and doped helium nanodroplets},\ }\href@noop {} {\bibfield  {journal}
  {\bibinfo  {journal} {Int. Rev. Phys. Chem.}\ }\textbf {\bibinfo {volume}
  {33}},\ \bibinfo {pages} {301} (\bibinfo {year} {2014})}\BibitemShut
  {NoStop}%
\bibitem [{\citenamefont {Radcliffe}\ \emph {et~al.}(2004)\citenamefont
  {Radcliffe}, \citenamefont {Przystawik}, \citenamefont {Diederich},
  \citenamefont {D\"oppner}, \citenamefont {Tiggesb\"aumker},\ and\
  \citenamefont {Meiwes-Broer}}]{Radcliffe:2004}%
  \BibitemOpen
  \bibfield  {author} {\bibinfo {author} {\bibfnamefont {P.}~\bibnamefont
  {Radcliffe}}, \bibinfo {author} {\bibfnamefont {A.}~\bibnamefont
  {Przystawik}}, \bibinfo {author} {\bibfnamefont {T.}~\bibnamefont
  {Diederich}}, \bibinfo {author} {\bibfnamefont {T.}~\bibnamefont
  {D\"oppner}}, \bibinfo {author} {\bibfnamefont {J.}~\bibnamefont
  {Tiggesb\"aumker}},\ and\ \bibinfo {author} {\bibfnamefont {K.-H.}\
  \bibnamefont {Meiwes-Broer}},\ }\bibfield  {title} {\bibinfo {title}
  {Excited-state relaxation of {A}g$_8$ clusters embedded in helium droplets},\
  }\href@noop {} {\bibfield  {journal} {\bibinfo  {journal} {Phys. Rev. Lett.}\
  }\textbf {\bibinfo {volume} {92}},\ \bibinfo {pages} {173403} (\bibinfo
  {year} {2004})}\BibitemShut {NoStop}%
\bibitem [{\citenamefont {Loginov}\ \emph {et~al.}(2005)\citenamefont
  {Loginov}, \citenamefont {Rossi},\ and\ \citenamefont
  {Drabbels}}]{Loginov:2005}%
  \BibitemOpen
  \bibfield  {author} {\bibinfo {author} {\bibfnamefont {E.}~\bibnamefont
  {Loginov}}, \bibinfo {author} {\bibfnamefont {D.}~\bibnamefont {Rossi}},\
  and\ \bibinfo {author} {\bibfnamefont {M.}~\bibnamefont {Drabbels}},\
  }\bibfield  {title} {\bibinfo {title} {Photoelectron spectroscopy of doped
  helium nanodroplets},\ }\href@noop {} {\bibfield  {journal} {\bibinfo
  {journal} {Phys. Rev. Lett.}\ }\textbf {\bibinfo {volume} {95}},\ \bibinfo
  {pages} {163401} (\bibinfo {year} {2005})}\BibitemShut {NoStop}%
\bibitem [{\citenamefont {Loginov}\ and\ \citenamefont
  {Drabbels}(2007)}]{Loginov:2007}%
  \BibitemOpen
  \bibfield  {author} {\bibinfo {author} {\bibfnamefont {E.}~\bibnamefont
  {Loginov}}\ and\ \bibinfo {author} {\bibfnamefont {M.}~\bibnamefont
  {Drabbels}},\ }\bibfield  {title} {\bibinfo {title} {Excited state dynamics
  of ag atoms in helium nanodroplets},\ }\href@noop {} {\bibfield  {journal}
  {\bibinfo  {journal} {J. Phys. Chem. A}\ }\textbf {\bibinfo {volume} {111}},\
  \bibinfo {pages} {7504} (\bibinfo {year} {2007})}\BibitemShut {NoStop}%
\bibitem [{\citenamefont {Loginov}\ \emph {et~al.}(2008)\citenamefont
  {Loginov}, \citenamefont {Braun},\ and\ \citenamefont
  {Drabbels}}]{Loginov:2008}%
  \BibitemOpen
  \bibfield  {author} {\bibinfo {author} {\bibfnamefont {E.}~\bibnamefont
  {Loginov}}, \bibinfo {author} {\bibfnamefont {A.}~\bibnamefont {Braun}},\
  and\ \bibinfo {author} {\bibfnamefont {M.}~\bibnamefont {Drabbels}},\
  }\bibfield  {title} {\bibinfo {title} {A new sensitive detection scheme for
  helium nanodroplet isolation spectroscopy: application to benzene},\
  }\href@noop {} {\bibfield  {journal} {\bibinfo  {journal} {Phys. Chem. Chem.
  Phys.}\ }\textbf {\bibinfo {volume} {10}},\ \bibinfo {pages} {6107} (\bibinfo
  {year} {2008})}\BibitemShut {NoStop}%
\bibitem [{\citenamefont {Fechner}\ \emph {et~al.}(2012)\citenamefont
  {Fechner}, \citenamefont {Gr{\"u}ner}, \citenamefont {Sieg}, \citenamefont
  {Callegari}, \citenamefont {Ancilotto}, \citenamefont {Stienkemeier},\ and\
  \citenamefont {Mudrich}}]{Fechner:2012}%
  \BibitemOpen
  \bibfield  {author} {\bibinfo {author} {\bibfnamefont {L.}~\bibnamefont
  {Fechner}}, \bibinfo {author} {\bibfnamefont {B.}~\bibnamefont {Gr{\"u}ner}},
  \bibinfo {author} {\bibfnamefont {A.}~\bibnamefont {Sieg}}, \bibinfo {author}
  {\bibfnamefont {C.}~\bibnamefont {Callegari}}, \bibinfo {author}
  {\bibfnamefont {F.}~\bibnamefont {Ancilotto}}, \bibinfo {author}
  {\bibfnamefont {F.}~\bibnamefont {Stienkemeier}},\ and\ \bibinfo {author}
  {\bibfnamefont {M.}~\bibnamefont {Mudrich}},\ }\bibfield  {title} {\bibinfo
  {title} {Photoionization and imaging spectroscopy of rubidium atoms attached
  to helium nanodroplets},\ }\href {https://doi.org/10.1039/C2CP22749E}
  {\bibfield  {journal} {\bibinfo  {journal} {Phys. Chem. Chem. Phys.}\
  }\textbf {\bibinfo {volume} {14}},\ \bibinfo {pages} {3843} (\bibinfo {year}
  {2012})}\BibitemShut {NoStop}%
\bibitem [{\citenamefont {Thaler}\ \emph {et~al.}(2018)\citenamefont {Thaler},
  \citenamefont {Ranftl}, \citenamefont {Heim}, \citenamefont {Cesnik},
  \citenamefont {Treiber}, \citenamefont {Meyer}, \citenamefont {Hauser},
  \citenamefont {Ernst},\ and\ \citenamefont {Koch}}]{Thaler:2018}%
  \BibitemOpen
  \bibfield  {author} {\bibinfo {author} {\bibfnamefont {B.}~\bibnamefont
  {Thaler}}, \bibinfo {author} {\bibfnamefont {S.}~\bibnamefont {Ranftl}},
  \bibinfo {author} {\bibfnamefont {P.}~\bibnamefont {Heim}}, \bibinfo {author}
  {\bibfnamefont {S.}~\bibnamefont {Cesnik}}, \bibinfo {author} {\bibfnamefont
  {L.}~\bibnamefont {Treiber}}, \bibinfo {author} {\bibfnamefont
  {R.}~\bibnamefont {Meyer}}, \bibinfo {author} {\bibfnamefont {A.~W.}\
  \bibnamefont {Hauser}}, \bibinfo {author} {\bibfnamefont {W.~E.}\
  \bibnamefont {Ernst}},\ and\ \bibinfo {author} {\bibfnamefont
  {M.}~\bibnamefont {Koch}},\ }\bibfield  {title} {\bibinfo {title}
  {Femtosecond photoexcitation dynamics inside a quantum solvent},\ }\href@noop
  {} {\bibfield  {journal} {\bibinfo  {journal} {Nature communications}\
  }\textbf {\bibinfo {volume} {9}},\ \bibinfo {pages} {1} (\bibinfo {year}
  {2018})}\BibitemShut {NoStop}%
\bibitem [{\citenamefont {Dozmorov}\ \emph {et~al.}(2018)\citenamefont
  {Dozmorov}, \citenamefont {Baklanov}, \citenamefont {von Vangerow},
  \citenamefont {Stienkemeier}, \citenamefont {Fordyce},\ and\ \citenamefont
  {Mudrich}}]{Dozmorov:2018}%
  \BibitemOpen
  \bibfield  {author} {\bibinfo {author} {\bibfnamefont {N.}~\bibnamefont
  {Dozmorov}}, \bibinfo {author} {\bibfnamefont {A.}~\bibnamefont {Baklanov}},
  \bibinfo {author} {\bibfnamefont {J.}~\bibnamefont {von Vangerow}}, \bibinfo
  {author} {\bibfnamefont {F.}~\bibnamefont {Stienkemeier}}, \bibinfo {author}
  {\bibfnamefont {J.}~\bibnamefont {Fordyce}},\ and\ \bibinfo {author}
  {\bibfnamefont {M.}~\bibnamefont {Mudrich}},\ }\bibfield  {title} {\bibinfo
  {title} {Quantum dynamics of rb atoms desorbing off the surface of he
  nanodroplets},\ }\href@noop {} {\bibfield  {journal} {\bibinfo  {journal}
  {Phys. Rev. A}\ }\textbf {\bibinfo {volume} {98}},\ \bibinfo {pages} {043403}
  (\bibinfo {year} {2018})}\BibitemShut {NoStop}%
\bibitem [{\citenamefont {Cederbaum}\ \emph {et~al.}(1997)\citenamefont
  {Cederbaum}, \citenamefont {Zobeley},\ and\ \citenamefont
  {Tarantelli}}]{Cederbaum:1997}%
  \BibitemOpen
  \bibfield  {author} {\bibinfo {author} {\bibfnamefont {L.~S.}\ \bibnamefont
  {Cederbaum}}, \bibinfo {author} {\bibfnamefont {J.}~\bibnamefont {Zobeley}},\
  and\ \bibinfo {author} {\bibfnamefont {F.}~\bibnamefont {Tarantelli}},\
  }\bibfield  {title} {\bibinfo {title} {Giant intermolecular decay and
  fragmentation of clusters},\ }\href@noop {} {\bibfield  {journal} {\bibinfo
  {journal} {Phys. Rev. Lett.}\ }\textbf {\bibinfo {volume} {79}},\ \bibinfo
  {pages} {4778} (\bibinfo {year} {1997})}\BibitemShut {NoStop}%
\bibitem [{\citenamefont {Shcherbinin}\ \emph {et~al.}(2017)\citenamefont
  {Shcherbinin}, \citenamefont {LaForge}, \citenamefont {Sharma}, \citenamefont
  {Devetta}, \citenamefont {Richter}, \citenamefont {Moshammer}, \citenamefont
  {Pfeifer},\ and\ \citenamefont {Mudrich}}]{Shcherbinin:2017}%
  \BibitemOpen
  \bibfield  {author} {\bibinfo {author} {\bibfnamefont {M.}~\bibnamefont
  {Shcherbinin}}, \bibinfo {author} {\bibfnamefont {A.~C.}\ \bibnamefont
  {LaForge}}, \bibinfo {author} {\bibfnamefont {V.}~\bibnamefont {Sharma}},
  \bibinfo {author} {\bibfnamefont {M.}~\bibnamefont {Devetta}}, \bibinfo
  {author} {\bibfnamefont {R.}~\bibnamefont {Richter}}, \bibinfo {author}
  {\bibfnamefont {R.}~\bibnamefont {Moshammer}}, \bibinfo {author}
  {\bibfnamefont {T.}~\bibnamefont {Pfeifer}},\ and\ \bibinfo {author}
  {\bibfnamefont {M.}~\bibnamefont {Mudrich}},\ }\bibfield  {title} {\bibinfo
  {title} {Interatomic coulombic decay in helium nanodroplets},\ }\href@noop {}
  {\bibfield  {journal} {\bibinfo  {journal} {Phys. Rev. A}\ }\textbf {\bibinfo
  {volume} {96}},\ \bibinfo {pages} {013407} (\bibinfo {year}
  {2017})}\BibitemShut {NoStop}%
\bibitem [{\citenamefont {Kelbg}\ \emph {et~al.}(2019)\citenamefont {Kelbg},
  \citenamefont {Zabel}, \citenamefont {Krebs}, \citenamefont {Kazak},
  \citenamefont {Meiwes-Broer},\ and\ \citenamefont
  {Tiggesb{\"a}umker}}]{Tiggesbaumker:2019}%
  \BibitemOpen
  \bibfield  {author} {\bibinfo {author} {\bibfnamefont {M.}~\bibnamefont
  {Kelbg}}, \bibinfo {author} {\bibfnamefont {M.}~\bibnamefont {Zabel}},
  \bibinfo {author} {\bibfnamefont {B.}~\bibnamefont {Krebs}}, \bibinfo
  {author} {\bibfnamefont {L.}~\bibnamefont {Kazak}}, \bibinfo {author}
  {\bibfnamefont {K.-H.}\ \bibnamefont {Meiwes-Broer}},\ and\ \bibinfo {author}
  {\bibfnamefont {J.}~\bibnamefont {Tiggesb{\"a}umker}},\ }\bibfield  {title}
  {\bibinfo {title} {Auger emission from the coulomb explosion of helium
  nanoplasmas},\ }\href@noop {} {\bibfield  {journal} {\bibinfo  {journal} {J.
  Chem. Phys.}\ }\textbf {\bibinfo {volume} {150}},\ \bibinfo {pages} {204302}
  (\bibinfo {year} {2019})}\BibitemShut {NoStop}%
\bibitem [{\citenamefont {LaForge}\ \emph {et~al.}(2019)\citenamefont
  {LaForge}, \citenamefont {Shcherbinin}, \citenamefont {Stienkemeier},
  \citenamefont {Richter}, \citenamefont {Moshammer}, \citenamefont {Pfeifer},\
  and\ \citenamefont {Mudrich}}]{LaForge:2019}%
  \BibitemOpen
  \bibfield  {author} {\bibinfo {author} {\bibfnamefont {A.}~\bibnamefont
  {LaForge}}, \bibinfo {author} {\bibfnamefont {M.}~\bibnamefont
  {Shcherbinin}}, \bibinfo {author} {\bibfnamefont {F.}~\bibnamefont
  {Stienkemeier}}, \bibinfo {author} {\bibfnamefont {R.}~\bibnamefont
  {Richter}}, \bibinfo {author} {\bibfnamefont {R.}~\bibnamefont {Moshammer}},
  \bibinfo {author} {\bibfnamefont {T.}~\bibnamefont {Pfeifer}},\ and\ \bibinfo
  {author} {\bibfnamefont {M.}~\bibnamefont {Mudrich}},\ }\bibfield  {title}
  {\bibinfo {title} {Highly efficient double ionization of mixed alkali dimers
  by intermolecular coulombic decay},\ }\href@noop {} {\bibfield  {journal}
  {\bibinfo  {journal} {Nature Physics}\ }\textbf {\bibinfo {volume} {15}},\
  \bibinfo {pages} {247} (\bibinfo {year} {2019})}\BibitemShut {NoStop}%
\bibitem [{\citenamefont {Ben~Ltaief}\ \emph {et~al.}(2019)\citenamefont
  {Ben~Ltaief}, \citenamefont {Shcherbinin}, \citenamefont {Mandal},
  \citenamefont {Krishnan}, \citenamefont {LaForge}, \citenamefont {Richter},
  \citenamefont {Turchini}, \citenamefont {Zema}, \citenamefont {Pfeifer},
  \citenamefont {Fasshauer} \emph {et~al.}}]{Ltaief:2019}%
  \BibitemOpen
  \bibfield  {author} {\bibinfo {author} {\bibfnamefont {L.}~\bibnamefont
  {Ben~Ltaief}}, \bibinfo {author} {\bibfnamefont {M.}~\bibnamefont
  {Shcherbinin}}, \bibinfo {author} {\bibfnamefont {S.}~\bibnamefont {Mandal}},
  \bibinfo {author} {\bibfnamefont {S.}~\bibnamefont {Krishnan}}, \bibinfo
  {author} {\bibfnamefont {A.}~\bibnamefont {LaForge}}, \bibinfo {author}
  {\bibfnamefont {R.}~\bibnamefont {Richter}}, \bibinfo {author} {\bibfnamefont
  {S.}~\bibnamefont {Turchini}}, \bibinfo {author} {\bibfnamefont
  {N.}~\bibnamefont {Zema}}, \bibinfo {author} {\bibfnamefont {T.}~\bibnamefont
  {Pfeifer}}, \bibinfo {author} {\bibfnamefont {E.}~\bibnamefont {Fasshauer}},
  \emph {et~al.},\ }\bibfield  {title} {\bibinfo {title} {Charge exchange
  dominates long-range interatomic coulombic decay of excited metal-doped
  helium nanodroplets},\ }\href@noop {} {\bibfield  {journal} {\bibinfo
  {journal} {J. Phys. Chem. Lett.}\ }\textbf {\bibinfo {volume} {10}},\
  \bibinfo {pages} {6904} (\bibinfo {year} {2019})}\BibitemShut {NoStop}%
\bibitem [{\citenamefont {Zobeley}\ \emph {et~al.}(2001)\citenamefont
  {Zobeley}, \citenamefont {Santra},\ and\ \citenamefont
  {Cederbaum}}]{Zobeley:2001}%
  \BibitemOpen
  \bibfield  {author} {\bibinfo {author} {\bibfnamefont {J.}~\bibnamefont
  {Zobeley}}, \bibinfo {author} {\bibfnamefont {R.}~\bibnamefont {Santra}},\
  and\ \bibinfo {author} {\bibfnamefont {L.~S.}\ \bibnamefont {Cederbaum}},\
  }\bibfield  {title} {\bibinfo {title} {Electronic decay in weakly bound
  heteroclusters: Energy transfer versus electron transfer},\ }\href@noop {}
  {\bibfield  {journal} {\bibinfo  {journal} {J. Chem. Phys.}\ }\textbf
  {\bibinfo {volume} {115}},\ \bibinfo {pages} {5076} (\bibinfo {year}
  {2001})}\BibitemShut {NoStop}%
\bibitem [{\citenamefont {LaForge}\ \emph {et~al.}(2016)\citenamefont
  {LaForge}, \citenamefont {Stumpf}, \citenamefont {Gokhberg}, \citenamefont
  {von Vangerow}, \citenamefont {Stienkemeier}, \citenamefont {Kryzhevoi},
  \citenamefont {O'Keeffe}, \citenamefont {Ciavardini}, \citenamefont
  {Krishnan}, \citenamefont {Coreno}, \citenamefont {Prince}, \citenamefont
  {Richter}, \citenamefont {Moshammer}, \citenamefont {Pfeifer}, \citenamefont
  {Cederbaum},\ and\ \citenamefont {Mudrich}}]{LaForgePRL:2016}%
  \BibitemOpen
  \bibfield  {author} {\bibinfo {author} {\bibfnamefont {A.~C.}\ \bibnamefont
  {LaForge}}, \bibinfo {author} {\bibfnamefont {V.}~\bibnamefont {Stumpf}},
  \bibinfo {author} {\bibfnamefont {K.}~\bibnamefont {Gokhberg}}, \bibinfo
  {author} {\bibfnamefont {J.}~\bibnamefont {von Vangerow}}, \bibinfo {author}
  {\bibfnamefont {F.}~\bibnamefont {Stienkemeier}}, \bibinfo {author}
  {\bibfnamefont {N.~V.}\ \bibnamefont {Kryzhevoi}}, \bibinfo {author}
  {\bibfnamefont {P.}~\bibnamefont {O'Keeffe}}, \bibinfo {author}
  {\bibfnamefont {A.}~\bibnamefont {Ciavardini}}, \bibinfo {author}
  {\bibfnamefont {S.~R.}\ \bibnamefont {Krishnan}}, \bibinfo {author}
  {\bibfnamefont {M.}~\bibnamefont {Coreno}}, \bibinfo {author} {\bibfnamefont
  {K.~C.}\ \bibnamefont {Prince}}, \bibinfo {author} {\bibfnamefont
  {R.}~\bibnamefont {Richter}}, \bibinfo {author} {\bibfnamefont
  {R.}~\bibnamefont {Moshammer}}, \bibinfo {author} {\bibfnamefont
  {T.}~\bibnamefont {Pfeifer}}, \bibinfo {author} {\bibfnamefont {L.~S.}\
  \bibnamefont {Cederbaum}},\ and\ \bibinfo {author} {\bibfnamefont
  {M.}~\bibnamefont {Mudrich}},\ }\bibfield  {title} {\bibinfo {title}
  {Enhanced ionization of embedded clusters by electron-transfer-mediated decay
  in helium nanodroplets},\ }\href@noop {} {\bibfield  {journal} {\bibinfo
  {journal} {Phys. Rev. Lett.}\ }\textbf {\bibinfo {volume} {116}},\ \bibinfo
  {pages} {203001} (\bibinfo {year} {2016})}\BibitemShut {NoStop}%
\bibitem [{\citenamefont {Ltaief}\ \emph {et~al.}(2020)\citenamefont {Ltaief},
  \citenamefont {Shcherbinin}, \citenamefont {Krishnan}, \citenamefont
  {Richter}, \citenamefont {Pfeifer}, \citenamefont {Bauer}, \citenamefont
  {Ghosh}, \citenamefont {Mudrich}, \citenamefont {Gokhberg}, \citenamefont
  {Laforge} \emph {et~al.}}]{Ltaief:2020}%
  \BibitemOpen
  \bibfield  {author} {\bibinfo {author} {\bibfnamefont {L.~B.}\ \bibnamefont
  {Ltaief}}, \bibinfo {author} {\bibfnamefont {M.}~\bibnamefont {Shcherbinin}},
  \bibinfo {author} {\bibfnamefont {S.}~\bibnamefont {Krishnan}}, \bibinfo
  {author} {\bibfnamefont {R.}~\bibnamefont {Richter}}, \bibinfo {author}
  {\bibfnamefont {T.}~\bibnamefont {Pfeifer}}, \bibinfo {author} {\bibfnamefont
  {M.}~\bibnamefont {Bauer}}, \bibinfo {author} {\bibfnamefont
  {A.}~\bibnamefont {Ghosh}}, \bibinfo {author} {\bibfnamefont
  {M.}~\bibnamefont {Mudrich}}, \bibinfo {author} {\bibfnamefont
  {K.}~\bibnamefont {Gokhberg}}, \bibinfo {author} {\bibfnamefont
  {A.}~\bibnamefont {Laforge}}, \emph {et~al.},\ }\bibfield  {title} {\bibinfo
  {title} {Electron transfer mediated decay of alkali dimers attached to he
  nanodroplets},\ }\href@noop {} {\bibfield  {journal} {\bibinfo  {journal}
  {Phys. Chem. Chem. Phys.}\ } (\bibinfo {year} {2020})}\BibitemShut {NoStop}%
\bibitem [{\citenamefont {Fr\"{o}chtenicht}\ \emph {et~al.}(1996)\citenamefont
  {Fr\"{o}chtenicht}, \citenamefont {Henne}, \citenamefont {Toennies},
  \citenamefont {Ding}, \citenamefont {Fieber-Erdmann},\ and\ \citenamefont
  {Drewello}}]{Froechtenicht:1996}%
  \BibitemOpen
  \bibfield  {author} {\bibinfo {author} {\bibfnamefont {R.}~\bibnamefont
  {Fr\"{o}chtenicht}}, \bibinfo {author} {\bibfnamefont {U.}~\bibnamefont
  {Henne}}, \bibinfo {author} {\bibfnamefont {J.~P.}\ \bibnamefont {Toennies}},
  \bibinfo {author} {\bibfnamefont {A.}~\bibnamefont {Ding}}, \bibinfo {author}
  {\bibfnamefont {M.}~\bibnamefont {Fieber-Erdmann}},\ and\ \bibinfo {author}
  {\bibfnamefont {T.}~\bibnamefont {Drewello}},\ }\bibfield  {title} {\bibinfo
  {title} {The photoionization of large pure and doped helium droplets},\
  }\href@noop {} {\bibfield  {journal} {\bibinfo  {journal} {J. Chem. Phys.}\
  }\textbf {\bibinfo {volume} {104}},\ \bibinfo {pages} {2548} (\bibinfo {year}
  {1996})}\BibitemShut {NoStop}%
\bibitem [{\citenamefont {Wang}\ \emph {et~al.}(2008)\citenamefont {Wang},
  \citenamefont {Kornilov}, \citenamefont {Gessner}, \citenamefont {Kim},
  \citenamefont {Peterka},\ and\ \citenamefont {Neumark}}]{Wang:2008}%
  \BibitemOpen
  \bibfield  {author} {\bibinfo {author} {\bibfnamefont {C.~C.}\ \bibnamefont
  {Wang}}, \bibinfo {author} {\bibfnamefont {O.}~\bibnamefont {Kornilov}},
  \bibinfo {author} {\bibfnamefont {O.}~\bibnamefont {Gessner}}, \bibinfo
  {author} {\bibfnamefont {J.~H.}\ \bibnamefont {Kim}}, \bibinfo {author}
  {\bibfnamefont {D.~S.}\ \bibnamefont {Peterka}},\ and\ \bibinfo {author}
  {\bibfnamefont {D.~M.}\ \bibnamefont {Neumark}},\ }\bibfield  {title}
  {\bibinfo {title} {Photoelectron imaging of helium droplets doped with {X}e
  and {K}r atoms},\ }\href@noop {} {\bibfield  {journal} {\bibinfo  {journal}
  {J. Phys. Chem.}\ }\textbf {\bibinfo {volume} {112}},\ \bibinfo {pages}
  {9356} (\bibinfo {year} {2008})}\BibitemShut {NoStop}%
\bibitem [{\citenamefont {Buchta}\ \emph
  {et~al.}(2013{\natexlab{a}})\citenamefont {Buchta}, \citenamefont {Krishnan},
  \citenamefont {Brauer}, \citenamefont {Drabbels}, \citenamefont {O'Keeffe},
  \citenamefont {Devetta}, \citenamefont {Di~Fraia}, \citenamefont {Callegari},
  \citenamefont {Richter}, \citenamefont {Coreno}, \citenamefont {Prince},
  \citenamefont {Stienkemeier}, \citenamefont {Moshammer},\ and\ \citenamefont
  {Mudrich}}]{Buchta:2013}%
  \BibitemOpen
  \bibfield  {author} {\bibinfo {author} {\bibfnamefont {D.}~\bibnamefont
  {Buchta}}, \bibinfo {author} {\bibfnamefont {S.~R.}\ \bibnamefont
  {Krishnan}}, \bibinfo {author} {\bibfnamefont {N.~B.}\ \bibnamefont
  {Brauer}}, \bibinfo {author} {\bibfnamefont {M.}~\bibnamefont {Drabbels}},
  \bibinfo {author} {\bibfnamefont {P.}~\bibnamefont {O'Keeffe}}, \bibinfo
  {author} {\bibfnamefont {M.}~\bibnamefont {Devetta}}, \bibinfo {author}
  {\bibfnamefont {M.}~\bibnamefont {Di~Fraia}}, \bibinfo {author}
  {\bibfnamefont {C.}~\bibnamefont {Callegari}}, \bibinfo {author}
  {\bibfnamefont {R.}~\bibnamefont {Richter}}, \bibinfo {author} {\bibfnamefont
  {M.}~\bibnamefont {Coreno}}, \bibinfo {author} {\bibfnamefont {K.~C.}\
  \bibnamefont {Prince}}, \bibinfo {author} {\bibfnamefont {F.}~\bibnamefont
  {Stienkemeier}}, \bibinfo {author} {\bibfnamefont {R.}~\bibnamefont
  {Moshammer}},\ and\ \bibinfo {author} {\bibfnamefont {M.}~\bibnamefont
  {Mudrich}},\ }\bibfield  {title} {\bibinfo {title} {Charge transfer and
  penning ionization of dopants in or on helium nanodroplets exposed to euv
  radiation},\ }\href {https://doi.org/10.1021/jp401424w} {\bibfield  {journal}
  {\bibinfo  {journal} {J. Phys. Chem. A}\ }\textbf {\bibinfo {volume} {117}},\
  \bibinfo {pages} {4394} (\bibinfo {year} {2013}{\natexlab{a}})}\BibitemShut
  {NoStop}%
\bibitem [{\citenamefont {Shcherbinin}\ \emph {et~al.}(2018)\citenamefont
  {Shcherbinin}, \citenamefont {LaForge}, \citenamefont {Hanif}, \citenamefont
  {Richter},\ and\ \citenamefont {Mudrich}}]{Shcherbinin:2018}%
  \BibitemOpen
  \bibfield  {author} {\bibinfo {author} {\bibfnamefont {M.}~\bibnamefont
  {Shcherbinin}}, \bibinfo {author} {\bibfnamefont {A.~C.}\ \bibnamefont
  {LaForge}}, \bibinfo {author} {\bibfnamefont {M.}~\bibnamefont {Hanif}},
  \bibinfo {author} {\bibfnamefont {R.}~\bibnamefont {Richter}},\ and\ \bibinfo
  {author} {\bibfnamefont {M.}~\bibnamefont {Mudrich}},\ }\bibfield  {title}
  {\bibinfo {title} {Penning ionization of acene molecules by helium
  nanodroplets},\ }\href@noop {} {\bibfield  {journal} {\bibinfo  {journal} {J.
  Phys. Chem. A}\ }\textbf {\bibinfo {volume} {122}},\ \bibinfo {pages} {1855}
  (\bibinfo {year} {2018})}\BibitemShut {NoStop}%
\bibitem [{\citenamefont {Scheidemann}\ \emph {et~al.}(1993)\citenamefont
  {Scheidemann}, \citenamefont {Schilling},\ and\ \citenamefont
  {Toennies}}]{Scheidemann:1993}%
  \BibitemOpen
  \bibfield  {author} {\bibinfo {author} {\bibfnamefont {A.}~\bibnamefont
  {Scheidemann}}, \bibinfo {author} {\bibfnamefont {B.}~\bibnamefont
  {Schilling}},\ and\ \bibinfo {author} {\bibfnamefont {J.~P.}\ \bibnamefont
  {Toennies}},\ }\bibfield  {title} {\bibinfo {title} {Anomalies in the
  reactions of he+ with sf6 embedded in large helium-4 clusters},\ }\href@noop
  {} {\bibfield  {journal} {\bibinfo  {journal} {J. Chem. Phys.}\ }\textbf
  {\bibinfo {volume} {97}},\ \bibinfo {pages} {2128} (\bibinfo {year}
  {1993})}\BibitemShut {NoStop}%
\bibitem [{\citenamefont {Mudrich}\ \emph {et~al.}(2020)\citenamefont
  {Mudrich}, \citenamefont {LaForge}, \citenamefont {Ciavardini}, \citenamefont
  {O’Keeffe}, \citenamefont {Callegari}, \citenamefont {Coreno},
  \citenamefont {Demidovich}, \citenamefont {Devetta}, \citenamefont
  {Di~Fraia}, \citenamefont {Drabbels} \emph {et~al.}}]{Mudrich:2020}%
  \BibitemOpen
  \bibfield  {author} {\bibinfo {author} {\bibfnamefont {M.}~\bibnamefont
  {Mudrich}}, \bibinfo {author} {\bibfnamefont {A.}~\bibnamefont {LaForge}},
  \bibinfo {author} {\bibfnamefont {A.}~\bibnamefont {Ciavardini}}, \bibinfo
  {author} {\bibfnamefont {P.}~\bibnamefont {O’Keeffe}}, \bibinfo {author}
  {\bibfnamefont {C.}~\bibnamefont {Callegari}}, \bibinfo {author}
  {\bibfnamefont {M.}~\bibnamefont {Coreno}}, \bibinfo {author} {\bibfnamefont
  {A.}~\bibnamefont {Demidovich}}, \bibinfo {author} {\bibfnamefont
  {M.}~\bibnamefont {Devetta}}, \bibinfo {author} {\bibfnamefont
  {M.}~\bibnamefont {Di~Fraia}}, \bibinfo {author} {\bibfnamefont
  {M.}~\bibnamefont {Drabbels}}, \emph {et~al.},\ }\bibfield  {title} {\bibinfo
  {title} {Ultrafast relaxation of photoexcited superfluid he nanodroplets},\
  }\href@noop {} {\bibfield  {journal} {\bibinfo  {journal} {Nature
  Communications}\ }\textbf {\bibinfo {volume} {11}} (\bibinfo {year}
  {2020})}\BibitemShut {NoStop}%
\bibitem [{\citenamefont {Lehmann}\ and\ \citenamefont
  {Northby}(1999)}]{Lehmann:1999}%
  \BibitemOpen
  \bibfield  {author} {\bibinfo {author} {\bibfnamefont {K.~K.}\ \bibnamefont
  {Lehmann}}\ and\ \bibinfo {author} {\bibfnamefont {J.~A.}\ \bibnamefont
  {Northby}},\ }\bibfield  {title} {\bibinfo {title} {Potential of an ionic
  impurity in a large he-4 cluster},\ }\href@noop {} {\bibfield  {journal}
  {\bibinfo  {journal} {Mol. Phys.}\ }\textbf {\bibinfo {volume} {97}},\
  \bibinfo {pages} {639} (\bibinfo {year} {1999})}\BibitemShut {NoStop}%
\bibitem [{\citenamefont {Mandal}\ \emph {et~al.}(2020)\citenamefont {Mandal},
  \citenamefont {Gopal}, \citenamefont {Shcherbinin}, \citenamefont {D'Elia},
  \citenamefont {Srinivas}, \citenamefont {Richter}, \citenamefont {Coreno},
  \citenamefont {Bapat}, \citenamefont {Mudrich}, \citenamefont {Krishnan},\
  and\ \citenamefont {Sharma}}]{Mandal:2020}%
  \BibitemOpen
  \bibfield  {author} {\bibinfo {author} {\bibfnamefont {S.}~\bibnamefont
  {Mandal}}, \bibinfo {author} {\bibfnamefont {R.}~\bibnamefont {Gopal}},
  \bibinfo {author} {\bibfnamefont {M.}~\bibnamefont {Shcherbinin}}, \bibinfo
  {author} {\bibfnamefont {A.}~\bibnamefont {D'Elia}}, \bibinfo {author}
  {\bibfnamefont {H.}~\bibnamefont {Srinivas}}, \bibinfo {author}
  {\bibfnamefont {R.}~\bibnamefont {Richter}}, \bibinfo {author} {\bibfnamefont
  {M.}~\bibnamefont {Coreno}}, \bibinfo {author} {\bibfnamefont
  {B.}~\bibnamefont {Bapat}}, \bibinfo {author} {\bibfnamefont
  {M.}~\bibnamefont {Mudrich}}, \bibinfo {author} {\bibfnamefont
  {S.}~\bibnamefont {Krishnan}},\ and\ \bibinfo {author} {\bibfnamefont
  {V.}~\bibnamefont {Sharma}},\ }\bibfield  {title} {\bibinfo {title} {Penning
  spectroscopy and structure of acetylene oligomers in he nanodroplets},\
  }\href@noop {} {\bibfield  {journal} {\bibinfo  {journal} {Phys. Chem. Chem.
  Phys.}\ } (\bibinfo {year} {2020})}\BibitemShut {NoStop}%
\bibitem [{\citenamefont {Buchta}\ \emph
  {et~al.}(2013{\natexlab{b}})\citenamefont {Buchta}, \citenamefont {Krishnan},
  \citenamefont {Brauer}, \citenamefont {Drabbels}, \citenamefont {O'Keeffe},
  \citenamefont {Devetta}, \citenamefont {Di~Fraia}, \citenamefont {Callegari},
  \citenamefont {Richter}, \citenamefont {Coreno}, \citenamefont {Prince},
  \citenamefont {Stienkemeier}, \citenamefont {Ullrich}, \citenamefont
  {Moshammer},\ and\ \citenamefont {Mudrich}}]{BuchtaJCP:2013}%
  \BibitemOpen
  \bibfield  {author} {\bibinfo {author} {\bibfnamefont {D.}~\bibnamefont
  {Buchta}}, \bibinfo {author} {\bibfnamefont {S.~R.}\ \bibnamefont
  {Krishnan}}, \bibinfo {author} {\bibfnamefont {N.~B.}\ \bibnamefont
  {Brauer}}, \bibinfo {author} {\bibfnamefont {M.}~\bibnamefont {Drabbels}},
  \bibinfo {author} {\bibfnamefont {P.}~\bibnamefont {O'Keeffe}}, \bibinfo
  {author} {\bibfnamefont {M.}~\bibnamefont {Devetta}}, \bibinfo {author}
  {\bibfnamefont {M.}~\bibnamefont {Di~Fraia}}, \bibinfo {author}
  {\bibfnamefont {C.}~\bibnamefont {Callegari}}, \bibinfo {author}
  {\bibfnamefont {R.}~\bibnamefont {Richter}}, \bibinfo {author} {\bibfnamefont
  {M.}~\bibnamefont {Coreno}}, \bibinfo {author} {\bibfnamefont {K.~C.}\
  \bibnamefont {Prince}}, \bibinfo {author} {\bibfnamefont {F.}~\bibnamefont
  {Stienkemeier}}, \bibinfo {author} {\bibfnamefont {J.}~\bibnamefont
  {Ullrich}}, \bibinfo {author} {\bibfnamefont {R.}~\bibnamefont {Moshammer}},\
  and\ \bibinfo {author} {\bibfnamefont {M.}~\bibnamefont {Mudrich}},\
  }\bibfield  {title} {\bibinfo {title} {Extreme ultraviolet ionization of pure
  he nanodroplets: Mass-correlated photoelectron imaging, penning ionization,
  and electron energy-loss spectra},\ }\href@noop {} {\bibfield  {journal}
  {\bibinfo  {journal} {J. Chem. Phys.}\ }\textbf {\bibinfo {volume} {139}},\
  \bibinfo {pages} {084301} (\bibinfo {year} {2013}{\natexlab{b}})}\BibitemShut
  {NoStop}%
\bibitem [{\citenamefont {Dick}(2014)}]{Dick:2014}%
  \BibitemOpen
  \bibfield  {author} {\bibinfo {author} {\bibfnamefont {B.}~\bibnamefont
  {Dick}},\ }\bibfield  {title} {\bibinfo {title} {Inverting ion images without
  abel inversion: maximum entropy reconstruction of velocity maps},\
  }\href@noop {} {\bibfield  {journal} {\bibinfo  {journal} {Physical Chemistry
  Chemical Physics}\ }\textbf {\bibinfo {volume} {16}},\ \bibinfo {pages} {570}
  (\bibinfo {year} {2014})}\BibitemShut {NoStop}%
\bibitem [{\citenamefont {Joppien}\ \emph {et~al.}(1993)\citenamefont
  {Joppien}, \citenamefont {Karnbach},\ and\ \citenamefont
  {M{\"o}ller}}]{Joppien:1993}%
  \BibitemOpen
  \bibfield  {author} {\bibinfo {author} {\bibfnamefont {M.}~\bibnamefont
  {Joppien}}, \bibinfo {author} {\bibfnamefont {R.}~\bibnamefont {Karnbach}},\
  and\ \bibinfo {author} {\bibfnamefont {T.}~\bibnamefont {M{\"o}ller}},\
  }\bibfield  {title} {\bibinfo {title} {Electronic excitations in
  liquid-helium: The evolution from small clusters to large droplets},\
  }\href@noop {} {\bibfield  {journal} {\bibinfo  {journal} {Phys. Rev. Lett.}\
  }\textbf {\bibinfo {volume} {71}},\ \bibinfo {pages} {2654} (\bibinfo {year}
  {1993})}\BibitemShut {NoStop}%
\bibitem [{\citenamefont {der Lan}\ \emph {et~al.}(2011)\citenamefont {der
  Lan}, \citenamefont {Bartl}, \citenamefont {Leidlmair}, \citenamefont
  {Sch\"{o}bel}, \citenamefont {Jochum}, \citenamefont {Denifl}, \citenamefont
  {M\"{a}rk}, \citenamefont {Ellis},\ and\ \citenamefont {Scheier}}]{Lan:2011}%
  \BibitemOpen
  \bibfield  {author} {\bibinfo {author} {\bibfnamefont {L.~A.}\ \bibnamefont
  {der Lan}}, \bibinfo {author} {\bibfnamefont {P.}~\bibnamefont {Bartl}},
  \bibinfo {author} {\bibfnamefont {C.}~\bibnamefont {Leidlmair}}, \bibinfo
  {author} {\bibfnamefont {H.}~\bibnamefont {Sch\"{o}bel}}, \bibinfo {author}
  {\bibfnamefont {R.}~\bibnamefont {Jochum}}, \bibinfo {author} {\bibfnamefont
  {S.}~\bibnamefont {Denifl}}, \bibinfo {author} {\bibfnamefont {T.~D.}\
  \bibnamefont {M\"{a}rk}}, \bibinfo {author} {\bibfnamefont {A.~M.}\
  \bibnamefont {Ellis}},\ and\ \bibinfo {author} {\bibfnamefont
  {P.}~\bibnamefont {Scheier}},\ }\bibfield  {title} {\bibinfo {title} {The
  submersion of sodium clusters in helium nanodroplets: Identification of the
  surface $\rightarrow$ interior transition},\ }\href@noop {} {\bibfield
  {journal} {\bibinfo  {journal} {J. Chem. Phys.}\ }\textbf {\bibinfo {volume}
  {135}},\ \bibinfo {eid} {044309} (\bibinfo {year} {2011})}\BibitemShut
  {NoStop}%
\bibitem [{\citenamefont {An~der Lan}\ \emph {et~al.}(2012)\citenamefont
  {An~der Lan}, \citenamefont {Bartl}, \citenamefont {Leidlmair}, \citenamefont
  {Sch\"obel}, \citenamefont {Denifl}, \citenamefont {M\"ark}, \citenamefont
  {Ellis},\ and\ \citenamefont {Scheier}}]{Lan:2012}%
  \BibitemOpen
  \bibfield  {author} {\bibinfo {author} {\bibfnamefont {L.}~\bibnamefont
  {An~der Lan}}, \bibinfo {author} {\bibfnamefont {P.}~\bibnamefont {Bartl}},
  \bibinfo {author} {\bibfnamefont {C.}~\bibnamefont {Leidlmair}}, \bibinfo
  {author} {\bibfnamefont {H.}~\bibnamefont {Sch\"obel}}, \bibinfo {author}
  {\bibfnamefont {S.}~\bibnamefont {Denifl}}, \bibinfo {author} {\bibfnamefont
  {T.~D.}\ \bibnamefont {M\"ark}}, \bibinfo {author} {\bibfnamefont {A.~M.}\
  \bibnamefont {Ellis}},\ and\ \bibinfo {author} {\bibfnamefont
  {P.}~\bibnamefont {Scheier}},\ }\bibfield  {title} {\bibinfo {title}
  {Submersion of potassium clusters in helium nanodroplets},\ }\href@noop {}
  {\bibfield  {journal} {\bibinfo  {journal} {Phys. Rev. B}\ }\textbf {\bibinfo
  {volume} {85}},\ \bibinfo {pages} {115414} (\bibinfo {year}
  {2012})}\BibitemShut {NoStop}%
\bibitem [{\citenamefont {Ederer}(1964)}]{Ederer:1964}%
  \BibitemOpen
  \bibfield  {author} {\bibinfo {author} {\bibfnamefont {D.~L.}\ \bibnamefont
  {Ederer}},\ }\bibfield  {title} {\bibinfo {title} {Photoionization of the 4d
  electrons in xenon},\ }\href@noop {} {\bibfield  {journal} {\bibinfo
  {journal} {Phys. Rev. Lett.}\ }\textbf {\bibinfo {volume} {13}},\ \bibinfo
  {pages} {760} (\bibinfo {year} {1964})}\BibitemShut {NoStop}%
\bibitem [{\citenamefont {Becker}\ \emph {et~al.}(1989)\citenamefont {Becker},
  \citenamefont {Szostak}, \citenamefont {Kerkhoff}, \citenamefont {Kupsch},
  \citenamefont {Langer}, \citenamefont {Wehlitz}, \citenamefont {Yagishita},\
  and\ \citenamefont {Hayaishi}}]{Becker:1989}%
  \BibitemOpen
  \bibfield  {author} {\bibinfo {author} {\bibfnamefont {U.}~\bibnamefont
  {Becker}}, \bibinfo {author} {\bibfnamefont {D.}~\bibnamefont {Szostak}},
  \bibinfo {author} {\bibfnamefont {H.}~\bibnamefont {Kerkhoff}}, \bibinfo
  {author} {\bibfnamefont {M.}~\bibnamefont {Kupsch}}, \bibinfo {author}
  {\bibfnamefont {B.}~\bibnamefont {Langer}}, \bibinfo {author} {\bibfnamefont
  {R.}~\bibnamefont {Wehlitz}}, \bibinfo {author} {\bibfnamefont
  {A.}~\bibnamefont {Yagishita}},\ and\ \bibinfo {author} {\bibfnamefont
  {T.}~\bibnamefont {Hayaishi}},\ }\bibfield  {title} {\bibinfo {title}
  {Subshell photoionization of {X}e between 40 and 1000 ev},\ }\href@noop {}
  {\bibfield  {journal} {\bibinfo  {journal} {Phys. Rev. A}\ }\textbf {\bibinfo
  {volume} {39}},\ \bibinfo {pages} {3902} (\bibinfo {year}
  {1989})}\BibitemShut {NoStop}%
\bibitem [{\citenamefont {Samson}\ and\ \citenamefont
  {Stolte}(2002)}]{Samson:2002}%
  \BibitemOpen
  \bibfield  {author} {\bibinfo {author} {\bibfnamefont {J.}~\bibnamefont
  {Samson}}\ and\ \bibinfo {author} {\bibfnamefont {W.~C.}\ \bibnamefont
  {Stolte}},\ }\bibfield  {title} {\bibinfo {title} {Precision measurements of
  the total photoionization cross-sections of he, ne, ar, kr, and xe},\
  }\href@noop {} {\bibfield  {journal} {\bibinfo  {journal} {Journal of
  electron spectroscopy and related phenomena}\ }\textbf {\bibinfo {volume}
  {123}},\ \bibinfo {pages} {265} (\bibinfo {year} {2002})}\BibitemShut
  {NoStop}%
\bibitem [{\citenamefont {Callicoatt}\ \emph {et~al.}(1996)\citenamefont
  {Callicoatt}, \citenamefont {Mar}, \citenamefont {Apkarian},\ and\
  \citenamefont {Janda}}]{Callicoatt:1996}%
  \BibitemOpen
  \bibfield  {author} {\bibinfo {author} {\bibfnamefont {B.}~\bibnamefont
  {Callicoatt}}, \bibinfo {author} {\bibfnamefont {D.}~\bibnamefont {Mar}},
  \bibinfo {author} {\bibfnamefont {A.}~\bibnamefont {Apkarian}},\ and\
  \bibinfo {author} {\bibfnamefont {K.}~\bibnamefont {Janda}},\ }\bibfield
  {title} {\bibinfo {title} {Charge transfer within he clusters},\ }\href@noop
  {} {\bibfield  {journal} {\bibinfo  {journal} {J. Chem. Phys.}\ }\textbf
  {\bibinfo {volume} {105}},\ \bibinfo {pages} {7872} (\bibinfo {year}
  {1996})}\BibitemShut {NoStop}%
\bibitem [{\citenamefont {Ruchti}\ \emph {et~al.}(2000)\citenamefont {Ruchti},
  \citenamefont {Callicoatt},\ and\ \citenamefont {Janda}}]{Ruchti:2000}%
  \BibitemOpen
  \bibfield  {author} {\bibinfo {author} {\bibfnamefont {T.}~\bibnamefont
  {Ruchti}}, \bibinfo {author} {\bibfnamefont {B.~E.}\ \bibnamefont
  {Callicoatt}},\ and\ \bibinfo {author} {\bibfnamefont {K.~C.}\ \bibnamefont
  {Janda}},\ }\bibfield  {title} {\bibinfo {title} {Charge transfer and
  fragmentation of liquid helium droplets doped with xenon},\ }\href@noop {}
  {\bibfield  {journal} {\bibinfo  {journal} {Phys. Chem. Chem. Phys.}\
  }\textbf {\bibinfo {volume} {2}},\ \bibinfo {pages} {4075} (\bibinfo {year}
  {2000})}\BibitemShut {NoStop}%
\bibitem [{\citenamefont {Gnodtke}\ \emph {et~al.}(2009)\citenamefont
  {Gnodtke}, \citenamefont {Saalmann},\ and\ \citenamefont
  {Rost}}]{Gnodtke:2009}%
  \BibitemOpen
  \bibfield  {author} {\bibinfo {author} {\bibfnamefont {C.}~\bibnamefont
  {Gnodtke}}, \bibinfo {author} {\bibfnamefont {U.}~\bibnamefont {Saalmann}},\
  and\ \bibinfo {author} {\bibfnamefont {J.~M.}\ \bibnamefont {Rost}},\
  }\bibfield  {title} {\bibinfo {title} {Ionization and charge migration
  through strong internal fields in clusters exposed to intense x-ray pulses},\
  }\href@noop {} {\bibfield  {journal} {\bibinfo  {journal} {Phys. Rev. A}\
  }\textbf {\bibinfo {volume} {79}},\ \bibinfo {pages} {041201} (\bibinfo
  {year} {2009})}\BibitemShut {NoStop}%
\bibitem [{\citenamefont {Oelze}\ \emph {et~al.}(2017)\citenamefont {Oelze},
  \citenamefont {Sch{\"u}tte}, \citenamefont {M{\"u}ller}, \citenamefont
  {M{\"u}ller}, \citenamefont {Wieland}, \citenamefont {Fr{\"u}hling},
  \citenamefont {Drescher}, \citenamefont {Al-Shemmary}, \citenamefont {Golz},
  \citenamefont {Stojanovic} \emph {et~al.}}]{oelze:2017}%
  \BibitemOpen
  \bibfield  {author} {\bibinfo {author} {\bibfnamefont {T.}~\bibnamefont
  {Oelze}}, \bibinfo {author} {\bibfnamefont {B.}~\bibnamefont {Sch{\"u}tte}},
  \bibinfo {author} {\bibfnamefont {M.}~\bibnamefont {M{\"u}ller}}, \bibinfo
  {author} {\bibfnamefont {J.~P.}\ \bibnamefont {M{\"u}ller}}, \bibinfo
  {author} {\bibfnamefont {M.}~\bibnamefont {Wieland}}, \bibinfo {author}
  {\bibfnamefont {U.}~\bibnamefont {Fr{\"u}hling}}, \bibinfo {author}
  {\bibfnamefont {M.}~\bibnamefont {Drescher}}, \bibinfo {author}
  {\bibfnamefont {A.}~\bibnamefont {Al-Shemmary}}, \bibinfo {author}
  {\bibfnamefont {T.}~\bibnamefont {Golz}}, \bibinfo {author} {\bibfnamefont
  {N.}~\bibnamefont {Stojanovic}}, \emph {et~al.},\ }\bibfield  {title}
  {\bibinfo {title} {Correlated electronic decay in expanding clusters
  triggered by intense xuv pulses from a free-electron-laser},\ }\href@noop {}
  {\bibfield  {journal} {\bibinfo  {journal} {Scientific reports}\ }\textbf
  {\bibinfo {volume} {7}},\ \bibinfo {pages} {1} (\bibinfo {year}
  {2017})}\BibitemShut {NoStop}%
\bibitem [{\citenamefont {Kumagai}\ \emph {et~al.}(2018)\citenamefont
  {Kumagai}, \citenamefont {Fukuzawa}, \citenamefont {Motomura}, \citenamefont
  {Iablonskyi}, \citenamefont {Nagaya}, \citenamefont {Wada}, \citenamefont
  {Ito}, \citenamefont {Takanashi}, \citenamefont {Sakakibara}, \citenamefont
  {You} \emph {et~al.}}]{Kumagai:2018}%
  \BibitemOpen
  \bibfield  {author} {\bibinfo {author} {\bibfnamefont {Y.}~\bibnamefont
  {Kumagai}}, \bibinfo {author} {\bibfnamefont {H.}~\bibnamefont {Fukuzawa}},
  \bibinfo {author} {\bibfnamefont {K.}~\bibnamefont {Motomura}}, \bibinfo
  {author} {\bibfnamefont {D.}~\bibnamefont {Iablonskyi}}, \bibinfo {author}
  {\bibfnamefont {K.}~\bibnamefont {Nagaya}}, \bibinfo {author} {\bibfnamefont
  {S.-i.}\ \bibnamefont {Wada}}, \bibinfo {author} {\bibfnamefont
  {Y.}~\bibnamefont {Ito}}, \bibinfo {author} {\bibfnamefont {T.}~\bibnamefont
  {Takanashi}}, \bibinfo {author} {\bibfnamefont {Y.}~\bibnamefont
  {Sakakibara}}, \bibinfo {author} {\bibfnamefont {D.}~\bibnamefont {You}},
  \emph {et~al.},\ }\bibfield  {title} {\bibinfo {title} {Following the birth
  of a nanoplasma produced by an ultrashort hard-x-ray laser in xenon
  clusters},\ }\href@noop {} {\bibfield  {journal} {\bibinfo  {journal}
  {Physical Review X}\ }\textbf {\bibinfo {volume} {8}},\ \bibinfo {pages}
  {031034} (\bibinfo {year} {2018})}\BibitemShut {NoStop}%
\bibitem [{\citenamefont {Atkins}(1959)}]{Atkins:1959}%
  \BibitemOpen
  \bibfield  {author} {\bibinfo {author} {\bibfnamefont {K.}~\bibnamefont
  {Atkins}},\ }\bibfield  {title} {\bibinfo {title} {Ions in liquid helium},\
  }\href@noop {} {\bibfield  {journal} {\bibinfo  {journal} {Phys. Rev.}\
  }\textbf {\bibinfo {volume} {116}},\ \bibinfo {pages} {1339} (\bibinfo {year}
  {1959})}\BibitemShut {NoStop}%
\bibitem [{\citenamefont {Tang}\ and\ \citenamefont
  {Toennies}(2003)}]{tang:2003}%
  \BibitemOpen
  \bibfield  {author} {\bibinfo {author} {\bibfnamefont {K.}~\bibnamefont
  {Tang}}\ and\ \bibinfo {author} {\bibfnamefont {J.}~\bibnamefont
  {Toennies}},\ }\bibfield  {title} {\bibinfo {title} {The van der waals
  potentials between all the rare gas atoms from he to rn},\ }\href@noop {}
  {\bibfield  {journal} {\bibinfo  {journal} {The Journal of chemical physics}\
  }\textbf {\bibinfo {volume} {118}},\ \bibinfo {pages} {4976} (\bibinfo {year}
  {2003})}\BibitemShut {NoStop}%
\bibitem [{\citenamefont {Hermann}\ and\ \citenamefont
  {Schwerdtfeger}(2009)}]{hermann:2009}%
  \BibitemOpen
  \bibfield  {author} {\bibinfo {author} {\bibfnamefont {A.}~\bibnamefont
  {Hermann}}\ and\ \bibinfo {author} {\bibfnamefont {P.}~\bibnamefont
  {Schwerdtfeger}},\ }\bibfield  {title} {\bibinfo {title} {Complete basis set
  limit second-order m{\o}ller--plesset calculations for the fcc lattices of
  neon, argon, krypton, and xenon},\ }\href@noop {} {\bibfield  {journal}
  {\bibinfo  {journal} {The Journal of chemical physics}\ }\textbf {\bibinfo
  {volume} {131}},\ \bibinfo {pages} {244508} (\bibinfo {year}
  {2009})}\BibitemShut {NoStop}%
\bibitem [{\citenamefont {Smolarek}\ \emph {et~al.}(2010)\citenamefont
  {Smolarek}, \citenamefont {Brauer}, \citenamefont {Buma},\ and\ \citenamefont
  {Drabbels}}]{smolarek:2010}%
  \BibitemOpen
  \bibfield  {author} {\bibinfo {author} {\bibfnamefont {S.}~\bibnamefont
  {Smolarek}}, \bibinfo {author} {\bibfnamefont {N.~B.}\ \bibnamefont
  {Brauer}}, \bibinfo {author} {\bibfnamefont {W.~J.}\ \bibnamefont {Buma}},\
  and\ \bibinfo {author} {\bibfnamefont {M.}~\bibnamefont {Drabbels}},\
  }\bibfield  {title} {\bibinfo {title} {Ir spectroscopy of molecular ions by
  nonthermal ion ejection from helium nanodroplets},\ }\href@noop {} {\bibfield
   {journal} {\bibinfo  {journal} {Journal of the American Chemical Society}\
  }\textbf {\bibinfo {volume} {132}},\ \bibinfo {pages} {14086} (\bibinfo
  {year} {2010})}\BibitemShut {NoStop}%
\bibitem [{\citenamefont {Lablanquie}\ \emph {et~al.}(2002)\citenamefont
  {Lablanquie}, \citenamefont {Sheinerman}, \citenamefont {Penent},
  \citenamefont {Hall}, \citenamefont {Ahmad}, \citenamefont {Aoto},
  \citenamefont {Hikosaka},\ and\ \citenamefont {Ito}}]{lablanquie:2002}%
  \BibitemOpen
  \bibfield  {author} {\bibinfo {author} {\bibfnamefont {P.}~\bibnamefont
  {Lablanquie}}, \bibinfo {author} {\bibfnamefont {S.}~\bibnamefont
  {Sheinerman}}, \bibinfo {author} {\bibfnamefont {F.}~\bibnamefont {Penent}},
  \bibinfo {author} {\bibfnamefont {R.}~\bibnamefont {Hall}}, \bibinfo {author}
  {\bibfnamefont {M.}~\bibnamefont {Ahmad}}, \bibinfo {author} {\bibfnamefont
  {T.}~\bibnamefont {Aoto}}, \bibinfo {author} {\bibfnamefont {Y.}~\bibnamefont
  {Hikosaka}},\ and\ \bibinfo {author} {\bibfnamefont {K.}~\bibnamefont
  {Ito}},\ }\bibfield  {title} {\bibinfo {title} {Photoemission of threshold
  electrons in the vicinity of the xenon 4d hole: dynamics of auger decay},\
  }\href@noop {} {\bibfield  {journal} {\bibinfo  {journal} {J. Phys. B: At.
  Mol. Opt. Phys.}\ }\textbf {\bibinfo {volume} {35}},\ \bibinfo {pages} {3265}
  (\bibinfo {year} {2002})}\BibitemShut {NoStop}%
\bibitem [{\citenamefont {Morishita}\ \emph {et~al.}(2006)\citenamefont
  {Morishita}, \citenamefont {Liu}, \citenamefont {Saito}, \citenamefont
  {Lischke}, \citenamefont {Kato}, \citenamefont {Pr{\"u}mper}, \citenamefont
  {Oura}, \citenamefont {Yamaoka}, \citenamefont {Tamenori}, \citenamefont
  {Suzuki} \emph {et~al.}}]{morishita:2006}%
  \BibitemOpen
  \bibfield  {author} {\bibinfo {author} {\bibfnamefont {Y.}~\bibnamefont
  {Morishita}}, \bibinfo {author} {\bibfnamefont {X.-J.}\ \bibnamefont {Liu}},
  \bibinfo {author} {\bibfnamefont {N.}~\bibnamefont {Saito}}, \bibinfo
  {author} {\bibfnamefont {T.}~\bibnamefont {Lischke}}, \bibinfo {author}
  {\bibfnamefont {M.}~\bibnamefont {Kato}}, \bibinfo {author} {\bibfnamefont
  {G.}~\bibnamefont {Pr{\"u}mper}}, \bibinfo {author} {\bibfnamefont
  {M.}~\bibnamefont {Oura}}, \bibinfo {author} {\bibfnamefont {H.}~\bibnamefont
  {Yamaoka}}, \bibinfo {author} {\bibfnamefont {Y.}~\bibnamefont {Tamenori}},
  \bibinfo {author} {\bibfnamefont {I.}~\bibnamefont {Suzuki}}, \emph
  {et~al.},\ }\bibfield  {title} {\bibinfo {title} {Experimental evidence of
  interatomic coulombic decay from the auger final states in argon dimers},\
  }\href@noop {} {\bibfield  {journal} {\bibinfo  {journal} {Physical review
  letters}\ }\textbf {\bibinfo {volume} {96}},\ \bibinfo {pages} {243402}
  (\bibinfo {year} {2006})}\BibitemShut {NoStop}%
\bibitem [{\citenamefont {Ueda}\ \emph {et~al.}(2007)\citenamefont {Ueda},
  \citenamefont {Liu}, \citenamefont {Pr{\"u}mper}, \citenamefont {Fukuzawa},
  \citenamefont {Morishita},\ and\ \citenamefont {Saito}}]{ueda:2007}%
  \BibitemOpen
  \bibfield  {author} {\bibinfo {author} {\bibfnamefont {K.}~\bibnamefont
  {Ueda}}, \bibinfo {author} {\bibfnamefont {X.-J.}\ \bibnamefont {Liu}},
  \bibinfo {author} {\bibfnamefont {G.}~\bibnamefont {Pr{\"u}mper}}, \bibinfo
  {author} {\bibfnamefont {H.}~\bibnamefont {Fukuzawa}}, \bibinfo {author}
  {\bibfnamefont {Y.}~\bibnamefont {Morishita}},\ and\ \bibinfo {author}
  {\bibfnamefont {N.}~\bibnamefont {Saito}},\ }\bibfield  {title} {\bibinfo
  {title} {Electron--ion coincidence momentum spectroscopy: Its application to
  ar dimer interatomic decay},\ }\href@noop {} {\bibfield  {journal} {\bibinfo
  {journal} {Journal of electron spectroscopy and related phenomena}\ }\textbf
  {\bibinfo {volume} {155}},\ \bibinfo {pages} {113} (\bibinfo {year}
  {2007})}\BibitemShut {NoStop}%
\bibitem [{\citenamefont {Ueda}\ \emph {et~al.}(2008)\citenamefont {Ueda},
  \citenamefont {Fukuzawa}, \citenamefont {Liu}, \citenamefont {Sakai},
  \citenamefont {Pr{\"u}mper}, \citenamefont {Morishita}, \citenamefont
  {Saito}, \citenamefont {Suzuki}, \citenamefont {Nagaya}, \citenamefont
  {Iwayama} \emph {et~al.}}]{ueda:2008}%
  \BibitemOpen
  \bibfield  {author} {\bibinfo {author} {\bibfnamefont {K.}~\bibnamefont
  {Ueda}}, \bibinfo {author} {\bibfnamefont {H.}~\bibnamefont {Fukuzawa}},
  \bibinfo {author} {\bibfnamefont {X.-J.}\ \bibnamefont {Liu}}, \bibinfo
  {author} {\bibfnamefont {K.}~\bibnamefont {Sakai}}, \bibinfo {author}
  {\bibfnamefont {G.}~\bibnamefont {Pr{\"u}mper}}, \bibinfo {author}
  {\bibfnamefont {Y.}~\bibnamefont {Morishita}}, \bibinfo {author}
  {\bibfnamefont {N.}~\bibnamefont {Saito}}, \bibinfo {author} {\bibfnamefont
  {I.}~\bibnamefont {Suzuki}}, \bibinfo {author} {\bibfnamefont
  {K.}~\bibnamefont {Nagaya}}, \bibinfo {author} {\bibfnamefont
  {H.}~\bibnamefont {Iwayama}}, \emph {et~al.},\ }\bibfield  {title} {\bibinfo
  {title} {Interatomic coulombic decay following the auger decay: Experimental
  evidence in rare-gas dimers},\ }\href@noop {} {\bibfield  {journal} {\bibinfo
   {journal} {Journal of Electron Spectroscopy and Related Phenomena}\ }\textbf
  {\bibinfo {volume} {166}},\ \bibinfo {pages} {3} (\bibinfo {year}
  {2008})}\BibitemShut {NoStop}%
\bibitem [{\citenamefont {Sakai}\ \emph {et~al.}(2011)\citenamefont {Sakai},
  \citenamefont {Stoychev}, \citenamefont {Ouchi}, \citenamefont {Higuchi},
  \citenamefont {Sch{\"o}ffler}, \citenamefont {Mazza}, \citenamefont
  {Fukuzawa}, \citenamefont {Nagaya}, \citenamefont {Yao}, \citenamefont
  {Tamenori} \emph {et~al.}}]{sakai:2011}%
  \BibitemOpen
  \bibfield  {author} {\bibinfo {author} {\bibfnamefont {K.}~\bibnamefont
  {Sakai}}, \bibinfo {author} {\bibfnamefont {S.}~\bibnamefont {Stoychev}},
  \bibinfo {author} {\bibfnamefont {T.}~\bibnamefont {Ouchi}}, \bibinfo
  {author} {\bibfnamefont {I.}~\bibnamefont {Higuchi}}, \bibinfo {author}
  {\bibfnamefont {M.}~\bibnamefont {Sch{\"o}ffler}}, \bibinfo {author}
  {\bibfnamefont {T.}~\bibnamefont {Mazza}}, \bibinfo {author} {\bibfnamefont
  {H.}~\bibnamefont {Fukuzawa}}, \bibinfo {author} {\bibfnamefont
  {K.}~\bibnamefont {Nagaya}}, \bibinfo {author} {\bibfnamefont
  {M.}~\bibnamefont {Yao}}, \bibinfo {author} {\bibfnamefont {Y.}~\bibnamefont
  {Tamenori}}, \emph {et~al.},\ }\bibfield  {title} {\bibinfo {title}
  {Electron-transfer-mediated decay and interatomic coulombic decay from the
  triply ionized states in argon dimers},\ }\href@noop {} {\bibfield  {journal}
  {\bibinfo  {journal} {Physical review letters}\ }\textbf {\bibinfo {volume}
  {106}},\ \bibinfo {pages} {033401} (\bibinfo {year} {2011})}\BibitemShut
  {NoStop}%
\bibitem [{\citenamefont {Liu}\ \emph {et~al.}(2020)\citenamefont {Liu},
  \citenamefont {Koloren\ifmmode~\check{c}\else \v{c}\fi{}},\ and\
  \citenamefont {Gokhberg}}]{Liu:2020}%
  \BibitemOpen
  \bibfield  {author} {\bibinfo {author} {\bibfnamefont {L.}~\bibnamefont
  {Liu}}, \bibinfo {author} {\bibfnamefont {P.~c.~v.}\ \bibnamefont
  {Koloren\ifmmode~\check{c}\else \v{c}\fi{}}},\ and\ \bibinfo {author}
  {\bibfnamefont {K.}~\bibnamefont {Gokhberg}},\ }\bibfield  {title} {\bibinfo
  {title} {Efficiency of core-level interatomic coulombic decay in rare-gas
  dimers},\ }\href {https://doi.org/10.1103/PhysRevA.101.033402} {\bibfield
  {journal} {\bibinfo  {journal} {Phys. Rev. A}\ }\textbf {\bibinfo {volume}
  {101}},\ \bibinfo {pages} {033402} (\bibinfo {year} {2020})}\BibitemShut
  {NoStop}%
\bibitem [{\citenamefont {Hans}\ \emph {et~al.}(2020)\citenamefont {Hans},
  \citenamefont {K\"ustner-Wetekam}, \citenamefont {Schmidt}, \citenamefont
  {Ozga}, \citenamefont {Holzapfel}, \citenamefont {Otto}, \citenamefont
  {Zindel}, \citenamefont {Richter}, \citenamefont {Cederbaum}, \citenamefont
  {Ehresmann}, \citenamefont {Hergenhahn}, \citenamefont {Kryzhevoi},\ and\
  \citenamefont {Knie}}]{hans:2020}%
  \BibitemOpen
  \bibfield  {author} {\bibinfo {author} {\bibfnamefont {A.}~\bibnamefont
  {Hans}}, \bibinfo {author} {\bibfnamefont {C.}~\bibnamefont
  {K\"ustner-Wetekam}}, \bibinfo {author} {\bibfnamefont {P.}~\bibnamefont
  {Schmidt}}, \bibinfo {author} {\bibfnamefont {C.}~\bibnamefont {Ozga}},
  \bibinfo {author} {\bibfnamefont {X.}~\bibnamefont {Holzapfel}}, \bibinfo
  {author} {\bibfnamefont {H.}~\bibnamefont {Otto}}, \bibinfo {author}
  {\bibfnamefont {C.}~\bibnamefont {Zindel}}, \bibinfo {author} {\bibfnamefont
  {C.}~\bibnamefont {Richter}}, \bibinfo {author} {\bibfnamefont {L.~S.}\
  \bibnamefont {Cederbaum}}, \bibinfo {author} {\bibfnamefont {A.}~\bibnamefont
  {Ehresmann}}, \bibinfo {author} {\bibfnamefont {U.}~\bibnamefont
  {Hergenhahn}}, \bibinfo {author} {\bibfnamefont {N.~V.}\ \bibnamefont
  {Kryzhevoi}},\ and\ \bibinfo {author} {\bibfnamefont {A.}~\bibnamefont
  {Knie}},\ }\bibfield  {title} {\bibinfo {title} {Core-level interatomic
  coulombic decay in van der waals clusters},\ }\href
  {https://doi.org/10.1103/PhysRevResearch.2.012022} {\bibfield  {journal}
  {\bibinfo  {journal} {Phys. Rev. Research}\ }\textbf {\bibinfo {volume}
  {2}},\ \bibinfo {pages} {012022} (\bibinfo {year} {2020})}\BibitemShut
  {NoStop}%
\bibitem [{\citenamefont {Bj{\"o}rneholm}\ \emph {et~al.}(2009)\citenamefont
  {Bj{\"o}rneholm}, \citenamefont {{\"O}hrwall},\ and\ \citenamefont
  {Tchaplyguine}}]{bjorneholm:2009}%
  \BibitemOpen
  \bibfield  {author} {\bibinfo {author} {\bibfnamefont {O.}~\bibnamefont
  {Bj{\"o}rneholm}}, \bibinfo {author} {\bibfnamefont {G.}~\bibnamefont
  {{\"O}hrwall}},\ and\ \bibinfo {author} {\bibfnamefont {M.}~\bibnamefont
  {Tchaplyguine}},\ }\bibfield  {title} {\bibinfo {title} {Free clusters
  studied by core-level spectroscopies},\ }\href@noop {} {\bibfield  {journal}
  {\bibinfo  {journal} {Nucl. Instrum. Methods Phys. Res. A}\ }\textbf
  {\bibinfo {volume} {601}},\ \bibinfo {pages} {161} (\bibinfo {year}
  {2009})}\BibitemShut {NoStop}%
\bibitem [{\citenamefont {Tchaplyguine}\ \emph {et~al.}(2004)\citenamefont
  {Tchaplyguine}, \citenamefont {Marinho}, \citenamefont {Gisselbrecht},
  \citenamefont {Schulz}, \citenamefont {M{\aa}rtensson}, \citenamefont
  {Sorensen}, \citenamefont {Naves~de Brito}, \citenamefont {Feifel},
  \citenamefont {{\"O}hrwall}, \citenamefont {Lundwall} \emph
  {et~al.}}]{tchaplyguine:2004}%
  \BibitemOpen
  \bibfield  {author} {\bibinfo {author} {\bibfnamefont {M.}~\bibnamefont
  {Tchaplyguine}}, \bibinfo {author} {\bibfnamefont {R.}~\bibnamefont
  {Marinho}}, \bibinfo {author} {\bibfnamefont {M.}~\bibnamefont
  {Gisselbrecht}}, \bibinfo {author} {\bibfnamefont {J.}~\bibnamefont
  {Schulz}}, \bibinfo {author} {\bibfnamefont {N.}~\bibnamefont
  {M{\aa}rtensson}}, \bibinfo {author} {\bibfnamefont {S.}~\bibnamefont
  {Sorensen}}, \bibinfo {author} {\bibfnamefont {A.}~\bibnamefont {Naves~de
  Brito}}, \bibinfo {author} {\bibfnamefont {R.}~\bibnamefont {Feifel}},
  \bibinfo {author} {\bibfnamefont {G.}~\bibnamefont {{\"O}hrwall}}, \bibinfo
  {author} {\bibfnamefont {M.}~\bibnamefont {Lundwall}}, \emph {et~al.},\
  }\bibfield  {title} {\bibinfo {title} {The size of neutral free clusters as
  manifested in the relative bulk-to-surface intensity in core level
  photoelectron spectroscopy},\ }\href@noop {} {\bibfield  {journal} {\bibinfo
  {journal} {J. Chem. Phys.}\ }\textbf {\bibinfo {volume} {120}},\ \bibinfo
  {pages} {345} (\bibinfo {year} {2004})}\BibitemShut {NoStop}%
\bibitem [{\citenamefont {Higgins}\ \emph {et~al.}(1996)\citenamefont
  {Higgins}, \citenamefont {Callegari}, \citenamefont {Reho}, \citenamefont
  {Stienkemeier}, \citenamefont {Ernst}, \citenamefont {Lehmann}, \citenamefont
  {Gutowski},\ and\ \citenamefont {Scoles}}]{Higgins:1996}%
  \BibitemOpen
  \bibfield  {author} {\bibinfo {author} {\bibfnamefont {J.}~\bibnamefont
  {Higgins}}, \bibinfo {author} {\bibfnamefont {C.}~\bibnamefont {Callegari}},
  \bibinfo {author} {\bibfnamefont {J.}~\bibnamefont {Reho}}, \bibinfo {author}
  {\bibfnamefont {F.}~\bibnamefont {Stienkemeier}}, \bibinfo {author}
  {\bibfnamefont {W.~E.}\ \bibnamefont {Ernst}}, \bibinfo {author}
  {\bibfnamefont {K.~K.}\ \bibnamefont {Lehmann}}, \bibinfo {author}
  {\bibfnamefont {M.}~\bibnamefont {Gutowski}},\ and\ \bibinfo {author}
  {\bibfnamefont {G.}~\bibnamefont {Scoles}},\ }\bibfield  {title} {\bibinfo
  {title} {Photoinduced chemical dynamics of high-spin alkali trimers},\
  }\href@noop {} {\bibfield  {journal} {\bibinfo  {journal} {Science}\ }\textbf
  {\bibinfo {volume} {273}},\ \bibinfo {pages} {629} (\bibinfo {year}
  {1996})}\BibitemShut {NoStop}%
\bibitem [{\citenamefont {Nauta}\ and\ \citenamefont
  {Miller}(1999)}]{NautaScience:1999}%
  \BibitemOpen
  \bibfield  {author} {\bibinfo {author} {\bibfnamefont {K.}~\bibnamefont
  {Nauta}}\ and\ \bibinfo {author} {\bibfnamefont {R.~E.}\ \bibnamefont
  {Miller}},\ }\bibfield  {title} {\bibinfo {title} {Nonequilibrium
  self-assembly of long chains of polar molecules in superfluid helium},\
  }\href@noop {} {\bibfield  {journal} {\bibinfo  {journal} {Science}\ }\textbf
  {\bibinfo {volume} {283}},\ \bibinfo {pages} {1895} (\bibinfo {year}
  {1999})}\BibitemShut {NoStop}%
\bibitem [{\citenamefont {Nauta}\ and\ \citenamefont
  {Miller}(2000)}]{NautaScience:2000}%
  \BibitemOpen
  \bibfield  {author} {\bibinfo {author} {\bibfnamefont {K.}~\bibnamefont
  {Nauta}}\ and\ \bibinfo {author} {\bibfnamefont {R.}~\bibnamefont {Miller}},\
  }\bibfield  {title} {\bibinfo {title} {Formation of cyclic water hexamer in
  liquid helium: The smallest piece of ice},\ }\href@noop {} {\bibfield
  {journal} {\bibinfo  {journal} {Science}\ }\textbf {\bibinfo {volume}
  {287}},\ \bibinfo {pages} {293} (\bibinfo {year} {2000})}\BibitemShut
  {NoStop}%
\bibitem [{\citenamefont {Przystawik}\ \emph {et~al.}(2008)\citenamefont
  {Przystawik}, \citenamefont {G{\"o}de}, \citenamefont {D{\"o}ppner},
  \citenamefont {Tiggesb{\"a}umker},\ and\ \citenamefont
  {Meiwes-Broer}}]{Przystawik:2008}%
  \BibitemOpen
  \bibfield  {author} {\bibinfo {author} {\bibfnamefont {A.}~\bibnamefont
  {Przystawik}}, \bibinfo {author} {\bibfnamefont {S.}~\bibnamefont
  {G{\"o}de}}, \bibinfo {author} {\bibfnamefont {T.}~\bibnamefont
  {D{\"o}ppner}}, \bibinfo {author} {\bibfnamefont {J.}~\bibnamefont
  {Tiggesb{\"a}umker}},\ and\ \bibinfo {author} {\bibfnamefont {K.-H.}\
  \bibnamefont {Meiwes-Broer}},\ }\bibfield  {title} {\bibinfo {title}
  {Light-induced collapse of metastable magnesium complexes formed in helium
  nanodroplets},\ }\href {https://doi.org/10.1103/PhysRevA.78.021202}
  {\bibfield  {journal} {\bibinfo  {journal} {Phys. Rev. A}\ }\textbf {\bibinfo
  {volume} {78}},\ \bibinfo {pages} {021202} (\bibinfo {year}
  {2008})}\BibitemShut {NoStop}%
\bibitem [{\citenamefont {Schomas}\ \emph {et~al.}(2020)\citenamefont {Schomas}
  \emph {et~al.}}]{Schomas:2020}%
  \BibitemOpen
  \bibfield  {author} {\bibinfo {author} {\bibfnamefont {D.}~\bibnamefont
  {Schomas}} \emph {et~al.},\ }\bibfield  {title} {\bibinfo {title} {Ignition
  of a helium nanoplasma by x-ray multiple ionization of a heavy rare-gas
  core},\ }\href@noop {} {\bibfield  {journal} {\bibinfo  {journal} {arXiv
  preprint arXiv:2005.02944}\ } (\bibinfo {year} {2020})}\BibitemShut {NoStop}%
\bibitem [{\citenamefont {Wabnitz}\ \emph {et~al.}(2002)\citenamefont
  {Wabnitz}, \citenamefont {Bittner}, \citenamefont {De~Castro}, \citenamefont
  {D{\"o}hrmann}, \citenamefont {G{\"u}rtler}, \citenamefont {Laarmann},
  \citenamefont {Laasch}, \citenamefont {Schulz}, \citenamefont {Swiderski},
  \citenamefont {von Haeften} \emph {et~al.}}]{Wabnitz:2002}%
  \BibitemOpen
  \bibfield  {author} {\bibinfo {author} {\bibfnamefont {H.}~\bibnamefont
  {Wabnitz}}, \bibinfo {author} {\bibfnamefont {L.}~\bibnamefont {Bittner}},
  \bibinfo {author} {\bibfnamefont {A.}~\bibnamefont {De~Castro}}, \bibinfo
  {author} {\bibfnamefont {R.}~\bibnamefont {D{\"o}hrmann}}, \bibinfo {author}
  {\bibfnamefont {P.}~\bibnamefont {G{\"u}rtler}}, \bibinfo {author}
  {\bibfnamefont {T.}~\bibnamefont {Laarmann}}, \bibinfo {author}
  {\bibfnamefont {W.}~\bibnamefont {Laasch}}, \bibinfo {author} {\bibfnamefont
  {J.}~\bibnamefont {Schulz}}, \bibinfo {author} {\bibfnamefont
  {A.}~\bibnamefont {Swiderski}}, \bibinfo {author} {\bibfnamefont
  {K.}~\bibnamefont {von Haeften}}, \emph {et~al.},\ }\bibfield  {title}
  {\bibinfo {title} {Multiple ionization of atom clusters by intense soft
  x-rays from a free-electron laser},\ }\href@noop {} {\bibfield  {journal}
  {\bibinfo  {journal} {Nature}\ }\textbf {\bibinfo {volume} {420}},\ \bibinfo
  {pages} {482} (\bibinfo {year} {2002})}\BibitemShut {NoStop}%
\bibitem [{\citenamefont {Hoener}\ \emph {et~al.}()\citenamefont {Hoener},
  \citenamefont {Bostedt}, \citenamefont {Thomas}, \citenamefont {Landt},
  \citenamefont {Eremina}, \citenamefont {Wabnitz}, \citenamefont {Laarmann},
  \citenamefont {Treusch}, \citenamefont {De~Castro},\ and\ \citenamefont
  {M{\"o}ller}}]{Hoener:2008}%
  \BibitemOpen
  \bibfield  {author} {\bibinfo {author} {\bibfnamefont {M.}~\bibnamefont
  {Hoener}}, \bibinfo {author} {\bibfnamefont {C.}~\bibnamefont {Bostedt}},
  \bibinfo {author} {\bibfnamefont {H.}~\bibnamefont {Thomas}}, \bibinfo
  {author} {\bibfnamefont {L.}~\bibnamefont {Landt}}, \bibinfo {author}
  {\bibfnamefont {E.}~\bibnamefont {Eremina}}, \bibinfo {author} {\bibfnamefont
  {H.}~\bibnamefont {Wabnitz}}, \bibinfo {author} {\bibfnamefont
  {T.}~\bibnamefont {Laarmann}}, \bibinfo {author} {\bibfnamefont
  {R.}~\bibnamefont {Treusch}}, \bibinfo {author} {\bibfnamefont
  {A.}~\bibnamefont {De~Castro}},\ and\ \bibinfo {author} {\bibfnamefont
  {T.}~\bibnamefont {M{\"o}ller}},\ }\bibfield  {title} {\bibinfo {title}
  {Charge recombination in soft x-ray laser produced nanoplasmas},\ }\href@noop
  {} {\bibinfo  {journal} {J. Phys. B}\ }\BibitemShut {NoStop}%
\bibitem [{\citenamefont {Gorkhover}\ \emph {et~al.}(2016)\citenamefont
  {Gorkhover}, \citenamefont {Schorb}, \citenamefont {Coffee}, \citenamefont
  {Adolph}, \citenamefont {Foucar}, \citenamefont {Rupp}, \citenamefont
  {Aquila}, \citenamefont {Bozek}, \citenamefont {Epp}, \citenamefont {Erk}
  \emph {et~al.}}]{Gorkhover:2016}%
  \BibitemOpen
\bibfield  {journal} {  }\bibfield  {author} {\bibinfo {author} {\bibfnamefont
  {T.}~\bibnamefont {Gorkhover}}, \bibinfo {author} {\bibfnamefont
  {S.}~\bibnamefont {Schorb}}, \bibinfo {author} {\bibfnamefont
  {R.}~\bibnamefont {Coffee}}, \bibinfo {author} {\bibfnamefont
  {M.}~\bibnamefont {Adolph}}, \bibinfo {author} {\bibfnamefont
  {L.}~\bibnamefont {Foucar}}, \bibinfo {author} {\bibfnamefont
  {D.}~\bibnamefont {Rupp}}, \bibinfo {author} {\bibfnamefont {A.}~\bibnamefont
  {Aquila}}, \bibinfo {author} {\bibfnamefont {J.~D.}\ \bibnamefont {Bozek}},
  \bibinfo {author} {\bibfnamefont {S.~W.}\ \bibnamefont {Epp}}, \bibinfo
  {author} {\bibfnamefont {B.}~\bibnamefont {Erk}}, \emph {et~al.},\ }\bibfield
   {title} {\bibinfo {title} {Femtosecond and nanometre visualization of
  structural dynamics in superheated nanoparticles},\ }\href@noop {} {\bibfield
   {journal} {\bibinfo  {journal} {Nature photonics}\ }\textbf {\bibinfo
  {volume} {10}},\ \bibinfo {pages} {93} (\bibinfo {year} {2016})}\BibitemShut
  {NoStop}%
\end{thebibliography}%
%

\end{document}